\documentclass[12pt]{article}
\usepackage{amssymb}
\usepackage{amsmath}
\usepackage{amsfonts}
\usepackage{sw20elba}
\usepackage{makeidx}
\usepackage[hypertex, bookmarksnumbered=true]{hyperref}

\input{tcilatex}
\begin{document}

\author{Anatoli Babin and Alexander Figotin \\
Department of Mathematics\\
University of California at Irvine\\
Irvine, CA 92697-3875}
\title{Neoclassical Theory of Elementary Charges with Spin of 1/2}
\maketitle

\begin{abstract}
We advance here our neoclassical theory of elementary charges by integrating
into it the concept of spin of 1/2. The developed spinorial version of our
theory has many important features identical to those of the Dirac theory
such as the gyromagnetic ratio, expressions for currents including the spin
current, and antimatter states. In our theory the concepts of charge and
anticharge relate naturally to their "spin" in its rest frame in two
opposite directions. An important difference with the Dirac theory is that
both the charge and anticharge energies are positive whereas their
frequencies have opposite signs.
\end{abstract}

\section{Introduction}

In a series of papers including \cite{BF6}-\cite{BF9} we have developed a
neoclassical theory of electromagnetic (EM) interactions between elementary
charges without spin. One of our key motivations for introducing such a
theory was a desire to account for particle properties as well as for wave
phenomena in a single mathematically sound Lagrangian relativistic field
theory. In this theory all particle properties come out naturally from the
field equations as approximations. We have shown that the theory implies in
the non-relativistic limit: (i) the non-relativistic particle mechanics
governed by the Newton equations with the Lorentz forces and (ii) the
frequency spectrum for hydrogenic atoms. We have studied also in \cite{BF8}, 
\cite{BF9} relativistic aspects of the theory and have demonstrated that the
relativistic point mass equation is an approximation of the field equations
when the charge wave function is well localized, and derived the Einstein
energy-mass relation $E=M\mathrm{c}^{2}$ for the accelerated motion.

A primary goal of this paper is to integrate into our neoclassical theory of
elementary charges the concept of spin of 1/2. As in the cited above papers,
an elementary charge is not a charged mass point, but it is described by a
field distribution, and now we want to add to its properties an intrinsic
magnetic moment and a spin. We have accomplished that goal by constructing a
spinorial version of the mentioned above neoclassical theory. When
developing this theory we kept in mind that it has to incorporate in one
form or another some features of the Dirac theory of spin 1/2 particles that
are verified experimentally. To integrate the spin into our Lagrangian
relativistic field theory we used methods developed by D. Hestenes and other
authors, see \cite{Hes2-03}, \cite{HesNF}, \cite{DeSabbataDatta}, \cite%
{DorLas}, \cite{Snygg} and references therein. In particular, we used D.
Hestenes's "real" form of the Dirac equation based on the spacetime algebra
(STA), that is the Clifford algebra of the Minkowski vector space. The
geometric transparency of the STA combined with a rich multivector algebraic
structure was a decisive incentive for using it instead of the Dirac $\gamma 
$-matrices.

Since the spinorial version of our neoclassical theory is obtained by a
modification of its original spinless version it is useful to take a look at
its basic features. In our original spinless theory a single elementary
charge is described by a pair $\left( \psi ,A^{\mu }\right) $, where $\psi $
is its complex valued \emph{wave function} and $A^{\mu }=\left( \varphi ,%
\mathbf{A}\right) $ is its 4-vector \emph{elementary potential} with the
corresponding \emph{elementary EM field} defined by the familiar formula $%
F^{\mu \nu }=\partial ^{\mu }A^{\nu }-\partial ^{\nu }A^{\mu }$. \emph{An
elementary charge does not interact with itself electromagnetically. Its
wave function }$\psi $\emph{\ represents its matter properties and the
elementary potential }$A^{\mu }$\emph{\ mediates its EM interactions with
all other elementary charges. Importantly,\ (i) all internal forces of an
elementary charge are exclusively of non-electromagnetic origin; (ii) every
elementary charge is a source of its elementary EM field which represents
force exerted by this charge on any other charge but not upon itself.}

A system of any number of elementary charges is furnished with a
relativistic Lagrangian that yields EM interactions with the following
features: (i) elementary charges interact only through their elementary EM
potentials and fields; (ii) the field equations for the elementary EM fields
are exactly the Maxwell equations with proper conserved currents; (iii) the
wave functions evolution is governed by nonlinear Klein-Gordon equations;
(iv) EM force density is described exactly by its well known Lorentz
expression; (v) the Newton equations with the Lorentz forces hold
approximately when charges are well separated and move with non-relativistic
velocities; (vi) a free charge moves uniformly preserving up to the Lorentz
contraction its shape. Since an overwhelming number of EM phenomena are
explained within the classical EM theory by the Maxwell equations and the
Lorentz forces our neoclassical EM theory is equally successful in
explaining the same phenomena.

Particle-like states of elementary charges are recovered in our theory from
the original field concepts as localized states. A possibility for an
elementary charge to localize is facilitated by internal forces of
non-electromagnetic origin. These forces are introduced in the Lagrangian in
the form of a nonlinear term $G\left( \left\vert \psi \right\vert
^{2}\right) $ defined by the following expression 
\begin{equation}
G\left( s\right) =G_{a}\left( s\right) =-a^{-2}s\left[ \ln \left(
a^{3}\left\vert s\right\vert \right) +\ln \pi ^{3/2}+2\right] ,\quad -\infty
<s<\infty ,  \label{nelagG1}
\end{equation}%
where $a>0$ is the size parameter. The free charge then has Gaussian shape
and is of the size $a$. As we have shown in \cite{BF6} \emph{the specific
expression (\ref{nelagG1}) for the nonlinearity }$G\left( s\right) $\emph{\
is derived from a physically sound requirement that the Planck-Einstein
relation }$E=\hbar \omega $\emph{\ holds exactly in the non-relativistic
approximation to our fully relativistic theory}.

To give a flavor of the proposed here spinorial version of our neoclassical
theory, we can state that the Euler-Lagrange field equation governing the
motion of a free charge is a spinorial version of the nonlinear Klein-Gordon
equation, namely%
\begin{equation}
\mathcal{P}^{2}\psi -\left[ \kappa _{0}^{2}+G^{\prime }\left( \left\langle
\psi \tilde{\psi}\right\rangle \right) \right] \psi =0,\quad \kappa _{0}=%
\frac{m\mathrm{c}}{\chi },  \label{pGKG}
\end{equation}%
where $\mathcal{P}$ is a certain spinorial version of the momentum operator
similar to the same in the Dirac theory, $G$ is a nonlinearity defined by (%
\ref{nelagG1}) $\chi $ is a constant which is equal approximately to the
Planck constant. We show that \emph{the spinorial version of our theory
developed here has many important features identical to those of the Dirac
theory such as the gyromagnetic ratio, expressions for currents including
the spin current, and antimatter states}. We treat here mostly the case of a
free charge, since the difference with the spinless scalar case shows itself
already in this case. Detailed analysis of more complex cases including
charge in external electromagnetic field or\ systems of interacting charges
is left for future work.

While our theory has many features identical to the Dirac theory as pointed
out above, it differs significantly from the Dirac theory. The first \
significant difference is that, by its very design, our neoclassical theory
is a consistent relativistic Lagrangian field theory. This difference is
manifest in the treatment of the energy. The Dirac theory following to the
quantum mechanics (QM) framework essentially identifies the energy with the
frequency through the Planck-Einstein relation $E=\hbar \omega $ considered
to be fundamentally exact. In our neoclassical Lagrangian field theory the
energy-momentum density is constructed based on the system Lagrangian and
the Noether theorem and its relation to frequencies in relevant regimes is
non-trivial. In particular, in our theory the Planck-Einstein relation holds
only approximately as a non-relativistic approximation for time harmonic
states taking the form $E\approx \hbar \left\vert \omega \right\vert $.
Observe that it is the compliance of the Dirac theory with the foundations
of QM requires the identity $E=\hbar \omega $ resulting in unbounded
negative energy problem. Indeed, the QM evolution equation $\hbar \mathrm{i}%
\partial _{t}\psi =H\psi $ with $H$ being the energy operator is just an
operator form of the Planck-Einstein relation $E=\hbar \omega $. This
evolution equation requires naturally negative frequencies which then
according to $E=\hbar \omega $ have to be interpreted as negative energies.

The second significant difference is that our Lagrangian theory is free from
any "infinities" which constitute a known problem for the QM and the quantum
electrodynamics.

The third difference is that in our theory an elementary charge is an
extended object which can be localized whereas in the QM it is a point-like
object. As R. Feynman put it, \cite[p. 21-6]{Feynman III} : "The wave
function $\psi \left( \mathbf{r}\right) $ for an electron in an atom does
not, then, describe a smeared-out electron with a smooth charge density. The
electron is either here, or there, or somewhere else, but wherever it is, it
is a point charge." In our theory the localization property of an elementary
charge in relevant situations is provided by a nonlinear non-electromagnetic
self-interaction $G$ defined by (\ref{nelagG1}). As a result the free charge
spinor wave function in our theory is a plane wave modulated by a Gaussian
amplitude factor compared to a plane wave free charge solution in the Dirac
theory. In particular, the localization property of the neoclassical free
charge solutions allows to evaluate the conserved quantities by integration
whereas that is not possible in the case of plane waves. The presence of the
nonlinearity in our theory also invalidates to some degree the linear
superposition principle, whereas it is of fundamental importance in the QM.

The fourth important difference is that in our theory there is no
electromagnetic self-interaction for an elementary charge.

The paper is organized as follows. In Section \ref{sbasSTA} we provide basic
information on the STA needed to carry out computations. In Section \ref%
{sDiracSTA} we discuss important properties of the STA version of the Dirac
theory. In Section \ref{sneobas} we develop the spinorial version of our
neoclassical theory and in Section \ref{sneofree} we study properties of a
free charge. In Section \ref{sneointer} we consider the interpretation of
the neoclassical solutions and compare main features of the developed here
theory with those of the Dirac theory.

\section{Basics of Spacetime Algebra (STA)\label{sbasSTA}}

In this section we formulate the basic properties of the \emph{Spacetime
Algebra} (STA), for it is perfectly suited for our conceptual purposes as
well the computation. The spacetime algebra is a particular case of the 
\emph{Clifford algebra}. A general \emph{Clifford algebra ,} also called 
\emph{Geometric Algebra} (GA), is an associative algebra generated by an $n$%
-dimensional vector space $V$ over the set of real scalars $\mathbb{R}$
furnished with a symmetric quadratic form $g$. We denote such a Clifford
algebra by $\mathrm{Cl}\left( V,g\right) $ and call its elements \emph{%
multivectors} referring to elements of the generating linear space $V$ as
vectors. The \emph{Clifford product}, also called \emph{geometric product},
of any two multivectors $A$ and $B$ is denoted by juxtaposition, that is $AB$%
. The Clifford product is fundamentally determined by the requirement to
satisfy the following identity for any two vectors $a$ and $b$ from the
generating vector space $V$: 
\begin{equation}
ab+ba=2g\left( a,b\right) 1,  \label{vecg1}
\end{equation}%
where $1$ is the multiplicative identity which we often skip in notation.
The Clifford algebra is naturally furnished with \emph{inner} "$\cdot $" 
\emph{(interior, dot) product and outer (exterior, Grassmann) product} "$%
\wedge $" so that for any two vectors $a$ and $b$ in $V$%
\begin{equation}
a\cdot b=\frac{ab+ba}{2}=g\left( a,b\right) ,\quad a\wedge b=-b\wedge a=%
\frac{ab-ba}{2},  \label{vecg2}
\end{equation}%
implying 
\begin{equation}
ab=a\cdot b+a\wedge b.  \label{vecg3}
\end{equation}%
\emph{According to (\ref{vecg2}), the orthogonality of two vectors }$a$\emph{%
\ and }$b$\emph{\ in the Clifford algebra, that is }$a\cdot b=0$\emph{, has
an equivalent algebraic representation as the anticommutativity of }$a$\emph{%
\ and }$b$.

\emph{The Spacetime Algebra is the Clifford algebra based on the real
4-dimensional Minkowski space} $\mathbb{M}^{4}$ and it is denoted by $%
\mathrm{Cl}\left( 1,3\right) $, where $\left( 1,3\right) $ is the signature
of the Minkowski space metric. The 3-dimensional Euclidian space is denoted
by $\mathbb{R}^{3}$ and the corresponding to it Clifford Algebra is $\mathrm{%
Cl}\left( 3,0\right) $. In setting up the STA we follow to \cite{Hes2-03}, 
\cite{DorLas} and \cite{Snygg}. Though many properties of Clifford Algebras
hold universally across different dimensions and signatures, we formulate
them mostly for the case of our primary interest which is the Spacetime
Algebra $\mathrm{Cl}\left( 1,3\right) $. We want to stress that the concise
review of the STA presented here is not meant to be complete and/or
systematic presentation of the Clifford Algebras theory, but rather it is a
selection of important for our purposes properties of the STA. We want to
acknowledge the work done by D. Hestenes who pioneered and developed many
aspects of the STA and its applications to physics. There is number of
excellent presentations of Clifford Algebras and their applications to
physics written by D. Hestenes and his followers, see, for instance, \cite%
{HesSob}, \cite{HesNF}, \cite{Hes2-03}, \cite{DorLas}, \cite{Snygg}, \cite%
{DorFonMan}.

The standard model for the spacetime is the real Minkowski vector space $%
\mathbb{M}^{4}$ with the standard metric $g_{\mu \nu }$ defined by%
\begin{equation}
\left\{ g_{\mu \nu }\right\} =\left\{ g^{\mu \nu }\right\} ,\quad
g_{00}=1,\quad g_{jj}=-1,\quad j=1,2,3,\quad g_{\mu \nu }=0,\quad \mu \neq
\nu .  \label{stag0}
\end{equation}%
A basis for the STA can be generated by a standard frame $\left\{ \gamma
_{\mu }:\mu =0,1,2,3\right\} $ of orthonormal vectors, with a \emph{timelike
vector} $\gamma _{0}$ in the forward light cone, and $\gamma _{\mu }$ are
assumed to satisfy the following relations:%
\begin{equation}
\gamma _{\mu }\gamma _{\nu }+\gamma _{\nu }\gamma _{\mu }=2\gamma _{\mu
}\cdot \gamma _{\nu }=2g_{\mu \nu },  \label{stag1}
\end{equation}%
\begin{equation}
\gamma _{0}^{2}=1,\quad \gamma _{i}^{2}=-1,\quad \gamma _{0}\cdot \gamma
_{i}=0,\quad \gamma _{i}\cdot \gamma _{j}=-\delta _{ij},\quad i,j=1,2,3.
\label{stag2}
\end{equation}%
Notice that (\ref{stag1})-(\ref{stag2}) are the defining relations of the 
\emph{Dirac matrix algebra}. That explains our choice to denote an
orthonormal frame by $\left\{ \gamma _{\mu }\right\} $, but it must be
remembered that the $\left\{ \gamma _{\mu }\right\} $ are basis vectors and
not a set of matrices in "isospace".

To facilitate algebraic manipulations, it is convenient to introduce the
reciprocal frame $\left\{ \gamma ^{\mu }\right\} $ defined by the equations%
\begin{equation}
\gamma ^{\mu }=g^{\mu \nu }\gamma _{\nu },\qquad \gamma _{\mu }\cdot \gamma
^{\nu }=\delta _{\mu }^{\nu },  \label{stag2a}
\end{equation}%
with the summation convention understood. Observe that two different vectors 
$\gamma _{\mu }$ anticommute. Since we are in a space of mixed signature, we
distinguish between a \emph{frame} $\left\{ \gamma _{\mu }\right\} $ \emph{%
and its reciprocal} $\left\{ \gamma ^{\mu }\right\} $, namely%
\begin{equation}
\gamma ^{0}=\gamma _{0},\qquad \gamma ^{i}=-\gamma _{i},\qquad i=1,2,3.
\label{stag3}
\end{equation}%
Notice that, following to the common practice, we use Greek letters $\mu
,\nu ,\ldots $ for indices taking values $0,1,2,3$ and Latin letter $%
i,j,\ldots $ for indices taking values $1,2,3$. The $\gamma _{\mu }$
determine a unique \emph{right-handed unit pseudoscalar}%
\begin{equation}
\mathrm{I}=\gamma _{0}\gamma _{1}\gamma _{2}\gamma _{3},\qquad \mathrm{I}%
^{2}=-1,  \label{stag4}
\end{equation}%
that anticommutes with vectors $\gamma _{\mu }$%
\begin{equation}
\mathrm{I}\gamma _{\mu }=-\gamma _{\mu }\mathrm{I},\qquad \mu =0,1,2,3.
\label{stag5}
\end{equation}%
For any vector $a$ a frame $\left\{ \gamma _{\mu }\right\} $ determines a
set of \emph{rectangular coordinates}%
\begin{equation}
a=a^{\mu }\gamma _{\mu }=a^{0}\gamma _{0}+a^{i}\gamma _{i},\qquad \left\{
a^{\mu }\right\} =\left\{ a^{0},\mathbf{a}\right\} .  \label{stag6}
\end{equation}%
In particular, for any spacetime point $x$%
\begin{equation}
x=x^{\mu }\gamma _{\mu }=ct\gamma _{0}+x^{i}\gamma _{i},\qquad \left\{
x^{\mu }\right\} =\left\{ x^{0},\mathbf{x}\right\} =\left\{ ct,\mathbf{x}%
\right\} .  \label{stag6a}
\end{equation}%
The frame $\left\{ \gamma _{\mu }\right\} $ defines also an explicit basis
for this algebra as follows:%
\begin{equation}
\underset{\text{1 scalar }}{1},\quad \underset{\text{4 vectors }}{\left\{
\gamma _{\mu }\right\} },\quad \underset{\text{6 bivectors }}{\left\{ \gamma
_{\mu }\wedge \gamma _{\nu }\right\} },\quad \underset{\text{4 trivectors }}{%
\left\{ \mathrm{I}\gamma _{\mu }\right\} },\quad \underset{\text{1
pseudoscalar }}{\left\{ \mathrm{I}\right\} },  \label{stag7}
\end{equation}%
where "$\wedge $" is the external (Grassman) product. This is the spacetime
algebra $\mathrm{Cl}\left( 1,3\right) $. The structure of this algebra tells
us practically all one needs to know about (flat) spacetime and the Lorentz
transformation group. A general element $M$ of the spacetime algebra is
called \emph{multivector} and can be written as%
\begin{equation}
M=\alpha +a+B+\mathrm{I}b+\mathrm{I}\beta ,  \label{stag8}
\end{equation}%
where $\alpha $ and $\beta $ are scalars, $a$ and $b$ are vectors, and $B$
is a bivector. The representation (\ref{stag8}) is a decomposition of $M$
into its $k$-vector parts (grades), and that can be expressed more
explicitly by putting it in the form 
\begin{eqnarray}
M &=&\sum_{0\leq k\leq 4}\left\langle M\right\rangle _{k},\quad \text{where }%
\left\langle M\right\rangle _{0}=\left\langle M\right\rangle =\alpha ,
\label{stag8a} \\
\left\langle M\right\rangle _{1} &=&a,\quad \left\langle M\right\rangle
_{2}=B,\quad \left\langle M\right\rangle _{3}=\mathrm{I}b,\quad \left\langle
M\right\rangle _{4}=\mathrm{I}\beta ,  \notag
\end{eqnarray}%
where the subscript $(k)$ means \textquotedblleft $k$-vector
part\textquotedblright . Notice the special notation $\left\langle
M\right\rangle =\left\langle M\right\rangle _{0}$ for the scalar part for
the multivector $M$. The space of $k$\emph{-vectors}, that is multivectors
of the grade $k$, is denoted by $\Lambda ^{k}$.

It is instructive to see the grade decomposition for the geometric product
of two multivectors $A_{r}\in \Lambda ^{r}$ and $B_{s}\in \Lambda ^{s}$, 
\cite[1.1]{HesSob}, \cite[2.2]{HesZie}, \cite[2.4.2]{Rodrigues Oliveira}%
\begin{gather}
A_{r}B_{s}=\left\langle A_{r}B_{s}\right\rangle _{\left\vert r-s\right\vert
}+\left\langle A_{r}B_{s}\right\rangle _{\left\vert r-s\right\vert
+2}+\cdots +\left\langle A_{r}B_{s}\right\rangle _{r+s}=  \label{stag8aa} \\
=\sum\nolimits_{k=0}^{m}\left\langle A_{r}B_{s}\right\rangle _{\left\vert
r-s\right\vert +2k},\text{ where }m=\frac{1}{2}\left( r+s-\left\vert
r-s\right\vert \right) ,  \notag
\end{gather}%
where it is understood that for any multivector $M$%
\begin{equation}
\left\langle M\right\rangle _{k}\equiv 0\text{ for any }k>4.  \label{stag8ab}
\end{equation}%
The inner (dot) "$\cdot $" and outer (Grassman) "$\wedge $" products are
defined first for homogeneous multivectors $A_{r}\in \Lambda ^{r}$ and $%
B_{s}\in \Lambda ^{s}$ by, \cite[1.1]{HesSob},%
\begin{gather}
A_{r}\cdot B_{s}=\left\langle A_{r}B_{s}\right\rangle _{\left\vert
r-s\right\vert },\text{ if }r,s>0;  \label{stag8b} \\
A_{r}\cdot A_{s}=0,\text{ if }r=0\text{ or }s=0;  \notag
\end{gather}%
\begin{equation}
A_{r}\cdot B_{s}=\left( -1\right) ^{r\left( s-r\right) }B_{s}\cdot A_{r}%
\text{ for }s\geq r;  \label{stag8ba}
\end{equation}%
\begin{equation}
A_{r}\wedge B_{s}=\left\langle A_{r}B_{s}\right\rangle _{r+s}=\left(
-1\right) ^{rs}B_{s}\wedge A_{r};  \label{stag8c}
\end{equation}%
with consequent extension by linearity to arbitrary multivectors $A$ and $B$%
. In particular, if $a$ is a vector and $A_{r}$ is multivector of the grade $%
r$, we have, \cite[1-1]{HesSob}, \cite[4.1.2]{DorLas}%
\begin{equation}
aA_{r}=a\cdot A_{r}+a\wedge A_{r},\qquad A_{r}a=A_{r}\cdot a+A_{r}\wedge a,
\label{stag8ca}
\end{equation}%
where%
\begin{gather}
a\cdot A_{r}=\left( -1\right) ^{r-1}A_{r}\cdot a=\frac{1}{2}\left(
aA_{r}-\left( -1\right) ^{r}A_{r}a\right) ,  \label{stag8cb} \\
a\wedge A_{r}=\left( -1\right) ^{r}A_{r}\wedge a=\frac{1}{2}\left(
aA_{r}+\left( -1\right) ^{r}A_{r}a\right) .  \notag
\end{gather}

The STA has a much richer structure than the algebra of complex numbers, and
it can be furnished with several natural conjugations (involutions)
operations, \cite[5.3]{DeSabbataDatta}, \cite[1.4.8]{Baylis}, \cite[3.1, 3.2]%
{Perw}. The most important of those is called \emph{reversion} (principal
anti-automorphism), and the \emph{reverse} $\tilde{M}$ of a general
multivector $M$ is defined by%
\begin{gather}
\tilde{M}=\alpha +a-B-\mathrm{I}b+\mathrm{I}\beta ,  \label{stag9} \\
\left\langle \tilde{M}\right\rangle _{k}=\widetilde{\left\langle
M\right\rangle _{k}}=\left( -1\right) ^{\frac{k\left( k-1\right) }{2}%
}\left\langle M\right\rangle _{k},\quad 0\leq k\leq 4.  \label{stag9a}
\end{gather}%
The reversion operation justifies its name since it reverses the order of
the multipliers: 
\begin{equation}
\left( MN\right) ^{\symbol{126}}=\tilde{N}\tilde{M}.  \label{stag10}
\end{equation}%
\emph{Grade involution} is a conjugation defined by%
\begin{equation}
\left\langle \hat{M}\right\rangle _{k}=\left( -1\right) ^{k}\left\langle
M\right\rangle _{k},\quad 0\leq k\leq 4.  \label{stag10b}
\end{equation}%
There is yet another \emph{Hermitian conjugation} also called \emph{relative
reversion} $M^{\dag }$ of a multivector $M$ defined by%
\begin{equation}
M^{\dag }=\gamma _{0}\tilde{M}\gamma _{0},  \label{stag10a}
\end{equation}%
and it corresponds to the Hermitian conjugation in the Dirac Algebra. Every
multivector then can be decomposed into $\gamma _{0}$-even and $\gamma _{0}$%
-odd components%
\begin{equation}
M=M_{\mathrm{e}}+M_{\mathrm{o}},\text{ where }M_{\mathrm{e}}=\frac{1}{2}%
\left( M^{\dag }+M\right) ,\quad M_{\mathrm{o}}=\frac{1}{2}\left( M^{\dag
}-M\right) ,  \label{stag10c}
\end{equation}%
and evidently%
\begin{equation}
M_{\mathrm{e}}^{\dag }=M_{\mathrm{e}},\quad M_{\mathrm{o}}^{\dag }=-M_{%
\mathrm{o}},\quad \tilde{M}^{\dag }=\widetilde{M^{\dag }}.  \label{stag10d}
\end{equation}

The grade structure (\ref{stag8}) of STA and the grade involution operator
defined by (\ref{stag10b}) provide for a natural decomposition of any
multivector $M$ into the sum of an \emph{even part} $M_{+}$ and an \emph{odd
part} $M_{-}$ as follows:%
\begin{equation}
M_{+}=\alpha +B+\mathrm{I}\beta ,\quad M_{-}=a+\mathrm{I}b,\quad M_{\pm }=%
\frac{1}{2}\left( M\pm \hat{M}\right) =\frac{1}{2}\left( M\mp \mathrm{I}M%
\mathrm{I}\right) .  \label{stag11}
\end{equation}%
Notice that the \emph{even and odd parts respectively commute and
anticommute with} $\mathrm{I}$\textrm{, }that is%
\begin{equation}
M_{+}\mathrm{I}=\mathrm{I}M_{+},\quad M_{-}\mathrm{I}=-M_{-}\mathrm{I}.
\label{stag12}
\end{equation}%
\emph{Importantly, the set of all even elements} $M_{+}$ of the STA $\mathrm{%
Cl}\left( 1,3\right) $ \emph{forms a Clifford algebra on its own, we denote
it by} $\mathrm{Cl}_{+}\left( 1,3\right) $. \emph{This even subalgebra }$%
\mathrm{Cl}_{+}\left( 1,3\right) $\emph{\ is isomorphic to the geometric
algebra (GA)} $\mathrm{Cl}\left( 3,0\right) $ \emph{of the three-dimensional
Euclidean space with multivectors of the form, \cite[1]{Hes-96}, \cite[VI]%
{Hes1-03},}%
\begin{equation}
N=\alpha +\mathrm{I}\beta +a+\mathrm{I}b\in \mathrm{Cl}\left( 3,0\right) ,
\label{stag13}
\end{equation}%
where $\alpha $ and $\beta $ are scalars, $a$ and $b$ are vectors and $%
\mathrm{I}$ is the unit pseudoscalar in $\mathrm{Cl}\left( 3,0\right) $. The
even subalgebra $\mathrm{Cl}_{+}\left( 1,3\right) $ is very important to the
STA version of the Dirac electron theory where it is the space of values of
the Dirac spinorial wave function.

Notice that the scalar part of $\left\langle M\right\rangle $ has the
following properties 
\begin{equation}
\left\langle M\right\rangle =\left\langle \tilde{M}\right\rangle ,\quad
\left\langle MN\right\rangle =\left\langle NM\right\rangle ,\quad
\left\langle \left\langle M\right\rangle _{k}\left\langle N\right\rangle
_{s}\right\rangle =0,\quad \text{if }k\neq s,  \label{scmn1}
\end{equation}%
where $M$ and $N$ are multivectors. The above equalities imply the following
identities involving Hermitian conjugation 
\begin{equation}
\left\langle MN^{\dag }\right\rangle =\left\langle N^{\dag }M\right\rangle
=\left\langle NM^{\dag }\right\rangle .  \label{scmn1H}
\end{equation}%
Based on the above we define first a \emph{scalar-valued }$\ast $\emph{%
-product} for any two arbitrary multivectors $A$ and $B$ by, \cite[1.1]%
{HesSob}, \cite[4.1.3]{DorLas}, \cite{DorstIP}, \cite[3.1.2]{DorFonMan}, 
\cite[3.2.3]{Perw}%
\begin{equation}
A\ast B=\left\langle AB\right\rangle =\sum_{0\leq k\leq 4}\left\langle
A_{\left( k\right) }B_{\left( k\right) }\right\rangle .  \label{scmn1a}
\end{equation}%
The above scalar $\ast $-product is symmetrical and reversible%
\begin{equation}
A\ast B=B\ast A=\tilde{A}\ast \tilde{B}=\tilde{B}\ast \tilde{A}.
\label{scmn1b}
\end{equation}%
Another (fiducial) \emph{scalar product} $A\centerdot B=\left\langle
A,B\right\rangle $ of two arbitrary multivectors $A$ and $B$ is defined by, 
\cite[3.4]{Snygg}, \cite[4.2.4]{Moya} 
\begin{gather}
A\centerdot B=\left\langle A,B\right\rangle =\tilde{A}\ast B=\left\langle 
\tilde{A}B\right\rangle =\left\langle A\tilde{B}\right\rangle  \label{scmn2}
\\
=\sum_{0\leq k\leq 4}\left\langle \widetilde{\left\langle A\right\rangle _{k}%
}\left\langle B\right\rangle _{k}\right\rangle =\sum_{0\leq k\leq 4}\left(
-1\right) ^{\frac{k\left( k-1\right) }{2}}\left\langle A_{\left( k\right)
}B_{\left( k\right) }\right\rangle .  \notag
\end{gather}%
Notice that we use the symbol "$\centerdot $" for the scalar product since
the "normal" dot symbol "$\cdot $" is already taken for the inner product.
Unfortunately, the symbols "$\ast $" and "$\cdot $" are used differently in
different texts and one has to pay attention when using those symbols. For
detailed and insightful analysis of relations between different products and
their geometric meaning see \cite{DorstIP}. The relations (\ref{scmn1})-(\ref%
{scmn3}) readily imply the following useful properties of the scalar products%
\begin{equation}
\left( AB\right) \ast C=A\ast \left( BC\right) =\left\langle ABC\right\rangle
\label{scmn2b}
\end{equation}%
\begin{eqnarray}
\left( AB\right) \centerdot C &=&B\centerdot \left( \tilde{A}C\right)
=A\centerdot \left( C\tilde{B}\right) ,  \label{scmn2a} \\
A\centerdot \left( BC\right) &=&\left( \tilde{B}A\right) \centerdot C=\left(
A\tilde{C}\right) \centerdot B.  \notag
\end{eqnarray}

A grade-$r$ multivector $A$ is called \emph{simple} or a \emph{blade} if it
is a product of $r$ anticommuting vectors, that is%
\begin{equation}
A=a_{1}\wedge a_{2}\cdots \wedge a_{r}\text{, where }a_{k}a_{j}=-a_{k}a_{j}%
\text{ for }k\neq j.  \label{scmn3}
\end{equation}%
Blades naturally correspond to subspaces, and they are instrumental to
establishing relations between geometric and algebraic properties. An
important property of every grade-$r$ blade $A_{r}$ is that it has the
inverse, \cite[1-1]{HesSob}, \cite[3.5.2]{DorFonMan}%
\begin{equation}
A_{r}^{-1}=\frac{\tilde{A}_{r}}{A_{r}\ast \tilde{A}_{r}}=\left( -1\right)
^{r\left( r-1\right) /2}\frac{A_{r}}{A_{r}\ast \tilde{A}_{r}}.
\label{scmn3a}
\end{equation}

In the case where $A_{r}$ and $B_{r}$ are simple $r$-vectors, the scalar
products (\ref{scmn1b}), (\ref{scmn2}) have the following representations
via the determinant, \cite[3.1.2]{DorFonMan}%
\begin{equation}
A_{r}\ast B_{r}=\left\langle A_{r}B_{r}\right\rangle =A_{r}\cdot B_{r}=\det 
\left[ 
\begin{array}{ccc}
\left\langle a_{1},b_{r}\right\rangle & \cdots & \left\langle
a_{1},b_{1}\right\rangle \\ 
\vdots & \ddots & \vdots \\ 
\left\langle a_{r},b_{r}\right\rangle & \cdots & \left\langle
a_{r},b_{1}\right\rangle%
\end{array}%
\right] ,\quad r>0,  \label{scmn4a}
\end{equation}%
\begin{equation}
A_{r}\centerdot B_{r}=A_{r}\ast \tilde{B}_{r}=\det \left[ 
\begin{array}{ccc}
\left\langle a_{1},b_{1}\right\rangle & \cdots & \left\langle
a_{1},b_{r}\right\rangle \\ 
\vdots & \ddots & \vdots \\ 
\left\langle a_{r},b_{1}\right\rangle & \cdots & \left\langle
a_{r},b_{r}\right\rangle%
\end{array}%
\right] ,\quad r>0,  \label{scmn4}
\end{equation}%
and%
\begin{equation}
A_{r}\ast B_{s}=0,\quad r\neq s;\quad a\ast b=ab,\text{ if }a\text{ and }b%
\text{ are scalars.}  \label{scmn4b}
\end{equation}%
Observe that in the case of the Clifford Algebra $\mathrm{Cl}\left(
3,0\right) $ of 3-dimensional Euclidian space, for any multivector $A\in 
\mathrm{Cl}\left( 3,0\right) $ the scalar product is positive, $A\centerdot
A=\left\langle \tilde{A}A\right\rangle \geq 0$; and that is the primary
motivation to define the scalar product by the formula (\ref{scmn1}). The
scalar product allows also to define a \emph{positive definite magnitude} $%
\left\vert M\right\vert $ for any multivector $M$ by%
\begin{equation}
\left\vert M\right\vert ^{2}=\left\vert \left\langle \tilde{M}M\right\rangle
\right\vert =\left\vert \left\langle M\tilde{M}\right\rangle \right\vert .
\label{scmn5}
\end{equation}%
Notice that in the case of vectors we always have%
\begin{equation}
A\cdot B=A\centerdot B=A\ast B\text{ if }A,B\in \Lambda ^{1}.  \label{scmn6}
\end{equation}

Being given a basis $\left\{ \gamma _{\mu }\right\} $ for $\mathcal{M}^{4}$,
we define a basis $\left\{ \mathbf{\sigma }_{k}\right\} $ for$\mathcal{\ }$%
the $3$-dimensional Euclidean space $\mathcal{P}^{3}$ by%
\begin{equation}
\mathbf{\sigma }_{k}=\gamma _{k}\wedge \gamma _{0}=\gamma _{k}\gamma _{0}=-%
\mathbf{\tilde{\sigma}}_{k}=-\mathbf{\sigma }^{k},\text{\qquad }\mathbf{%
\sigma }_{k}^{2}=1,\quad k=1,2,3,  \label{sigmg1}
\end{equation}%
\begin{equation}
\mathbf{\sigma }_{i}\mathbf{\sigma }_{j}=-\gamma _{i}\gamma _{j}=-\gamma
_{i}\wedge \gamma _{j}=\epsilon _{ijk}\mathrm{I}\mathbf{\sigma }_{k},\quad
i\neq j,\quad \mathbf{\sigma }_{1}\mathbf{\sigma }_{2}\mathbf{\sigma }_{3}=%
\mathrm{I},\quad \left( \mathrm{I}\mathbf{\sigma }_{k}\right) ^{2}=-1,
\label{sigmg2}
\end{equation}%
where $\epsilon _{ijk}$ is the \emph{alternating tensor}, also called \emph{%
Levi-Civita symbol}, defined by%
\begin{equation}
\epsilon _{ijk}=\left\{ 
\begin{tabular}{lll}
$1$ & if & $ijk$ is a cyclic permutation of $123,$ \\ 
$-1$ & if & $ijk$ is a anticyclic permutation of $123,$ \\ 
$0$ & if & otherwise.%
\end{tabular}%
\right. .  \label{sigmg2a}
\end{equation}%
Notice also that the following identities hold%
\begin{equation}
\gamma _{0}\mathbf{\sigma }_{k}=-\mathbf{\sigma }_{k}\gamma _{0},\quad
\gamma _{0}\mathrm{I}=-\mathrm{I}\gamma _{0},\quad \gamma _{0}\mathrm{I}%
\mathbf{\sigma }_{k}=\mathrm{I}\mathbf{\sigma }_{k}\gamma _{0},\quad \mathbf{%
\sigma }_{k}\mathrm{I}=\mathrm{I}\mathbf{\sigma }_{k},\quad k=1,2,3.
\label{sigmg3}
\end{equation}%
\begin{equation}
\mathbf{\sigma }_{i}\cdot \mathbf{\sigma }_{j}=\delta _{ij},\quad \frac{1}{2}%
\left( \mathbf{\sigma }_{i}\mathbf{\sigma }_{j}-\mathbf{\sigma }_{i}\mathbf{%
\sigma }_{j}\right) =\epsilon _{ijk}\mathrm{I}\mathbf{\sigma }_{k},\quad 
\frac{1}{2}\left( \mathrm{I}\mathbf{\sigma }_{i}\mathrm{I}\mathbf{\sigma }%
_{j}-\mathrm{I}\mathbf{\sigma }_{i}\mathrm{I}\mathbf{\sigma }_{j}\right)
=\epsilon _{ijk}\mathrm{I}\mathbf{\sigma }_{k}.  \label{sigmg4}
\end{equation}%
The bivectors $\mathbf{\sigma }_{k}$ are called \emph{relative vectors} and
they correspond to timelike planes. The \emph{relative vectors} $\mathbf{%
\sigma }_{k}$ \emph{generate the even subalgebra} $\mathrm{Cl}_{+}\left(
1,3\right) $ \emph{which is isomorphic to the geometric algebra (GA)} $%
\mathrm{Cl}\left( 3,0\right) $ \emph{of the three-dimensional Euclidean
space, \cite[3]{Hes-86}, \cite[1]{Hes-96}.} Relative bivectors $\mathrm{I}%
\mathbf{\sigma }_{k}$ according to (\ref{sigmg2}) are spacelike bivectors.

Observe that using $\mathrm{I}^{2}=-1$, we can recast the relations (\ref%
{sigmg2}) as 
\begin{equation}
\mathrm{I}\mathbf{\sigma }_{i}\mathrm{I}\mathbf{\sigma }_{j}=-\mathbf{\sigma 
}_{i}\mathbf{\sigma }_{j}=\gamma _{i}\gamma _{j}=\gamma _{i}\wedge \gamma
_{j}=-\epsilon _{ijk}\mathrm{I}\mathbf{\sigma }_{k},\qquad i\neq j,
\label{sigmg5}
\end{equation}%
implying that the span $\left\langle 1,\mathrm{I}\mathbf{\sigma }_{1},%
\mathrm{I}\mathbf{\sigma }_{2},\mathrm{I}\mathbf{\sigma }_{3}\right\rangle $
is a subalgebra $\mathsf{Q}$ which is isomorphic to the even subalgebra $%
\mathrm{Cl}_{+}\left( 3,0\right) $ of the geometric algebra $\mathrm{Cl}%
\left( 3,0\right) $ of the three-dimensional Euclidian space. Since $\mathrm{%
Cl}_{+}\left( 3,0\right) $ is isomorphic to the \emph{quaternion algebra}, 
\cite[2.3]{HesNF}, \cite[2.4.2]{DorLas}, \cite[6.1]{DeSabbataDatta}, the
subalgebra $\mathsf{Q}$ \emph{is also isomorphic to the quaternion algebra}
and we refer to it by that name, that is 
\begin{equation}
\mathsf{Q}=\left\langle 1,\mathrm{I}\mathbf{\sigma }_{1},\mathrm{I}\mathbf{%
\sigma }_{2},\mathrm{I}\mathbf{\sigma }_{3}\right\rangle \text{ is the
quaternion subalgebra.}  \label{sigmg6}
\end{equation}%
Notice that the quaternion subalgebra $\mathsf{Q}$ can also be characterized
as the one consisting of even multivectors which are also $\gamma _{0}$%
-even, that is 
\begin{equation}
\mathsf{Q}=\left\{ M\in \mathrm{Cl}_{+}\left( 3,0\right) :M^{\dag }=\gamma
_{0}\tilde{M}\gamma _{0}=M\right\} .  \label{sigmg7}
\end{equation}%
The quaternion subalgebra $\mathsf{Q}$ is very important to the STA version
of the Pauli electron theory where it is the space of values of the Pauli
spinorial wave function.

\section{The Dirac equation in STA\label{sDiracSTA}}

Since the Dirac theory has been very thoroughly analyzed and tested
experimentally, we would like to consider its STA version in sufficient
detail and compare it with developed here neoclassical theory. In addition
to that, the Dirac equation in the STA and its analysis provides us with a
number of valuable tools useful for our own constructions, and we consider
its important features in this section.

The STA version of Dirac spinor $\Psi $ is \emph{the wave function }$\psi $%
\emph{\ taking values in the even subalgebra} $\mathrm{Cl}_{+}\left(
1,3\right) $ of the Clifford algebra $\mathrm{Cl}\left( 1,3\right) $, and we
refer to it as \emph{Dirac spinor} or just \emph{spinor}. Notice that for
any $\psi $ from $\mathrm{Cl}_{+}\left( 1,3\right) $ we have $\psi \tilde{%
\psi}=\widetilde{\psi \tilde{\psi}}$ implying that this product is a linear
combination of the scalar and the pseudoscalar $\mathrm{I}$, that is 
\begin{equation}
\psi \tilde{\psi}=\tilde{\psi}\psi =\varrho \mathrm{e}^{\mathrm{I}\beta
}=\varrho \left( \cos \beta +\mathrm{I}\sin \beta \right) ,\text{ where }%
\varrho \geq 0\text{ and }\beta \text{ are scalars.}  \label{wfudisp1}
\end{equation}%
This leads to the following \emph{canonical Lorentz invariant decomposition}
which holds for every even multivector $\psi $, \cite{Hes-75}, \cite[VII.D]%
{Hes2-03}, \cite[8.2]{DorLas}, \cite[9.3]{DeSabbataDatta}, 
\begin{equation}
\psi =\varrho ^{\frac{1}{2}}\mathrm{e}^{\frac{\mathrm{I}\beta }{2}%
}R=R\varrho ^{\frac{1}{2}}\mathrm{e}^{\frac{\mathrm{I}\beta }{2}},\qquad R%
\tilde{R}=R\tilde{R}=1,  \label{wfudisp2}
\end{equation}%
where $\varrho >0$ and $\beta $ are scalars, and $R$ is the Lorentz rotor,
that is $x^{\prime }=Rx\tilde{R}$ is the Lorentz transformation. According
to D. Hestenes, the canonical decomposition (\ref{wfudisp1}) can be regarded
as an invariant decomposition of the Dirac wave function into a 2-parameter 
\emph{statistical factor} $\varrho ^{\frac{1}{2}}\mathrm{e}^{\frac{\mathrm{I}%
\beta }{2}}$ and a 6-parameter \emph{kinematical factor} $R$.

It is worth to point out that the identity (\ref{wfudisp1}) clearly shows
that though the reversion operation $\tilde{\psi}$ is analogous to the
complex conjugation for complex numbers, the even subalgebra $\mathrm{Cl}%
_{+}\left( 1,3\right) $ is a richer entity than the set of complex numbers
allowing $\psi \tilde{\psi}$ \emph{to be negative and not scalar valued}.

To introduce an STA form of the Dirac equation, we define first an STA
version of the \emph{Dirac operator} denoted sometimes by nabla dagger, \cite%
[2-1-2]{ItzZub}, \cite[3]{GreinerRQM}. We denote this STA version of the
Dirac operator by $\partial =\partial _{x}$. It is often called \emph{vector
derivative} with respect to vector $x$ and defined by, \cite{Hes1-03}, \cite[%
II]{Hes2-03}, \cite{HesSob}, \cite{DorLas}, 
\begin{equation}
\partial =\partial _{x}=\gamma ^{\mu }\partial _{\mu },\text{ where }%
\partial _{\mu }=\frac{\partial }{\partial x^{\mu }}.  \label{diracop1}
\end{equation}%
Notice that since $\partial $ is a vector, it may not commute with other
multivectors.

In the case of the Clifford algebra $\mathrm{Cl}\left( 3,0\right) $ of the
3-dimensional Euclidian space,\emph{\ the vector derivative} $\nabla $ is
defined by%
\begin{equation}
\nabla =\dsum\limits_{j=1}^{3}\mathbf{\sigma }_{j}\partial _{j},\text{ where 
}\partial _{j}=\frac{\partial }{\partial x^{j}},\text{ and }\mathbf{\sigma }%
_{j}\text{ is a basis of }\mathrm{Cl}\left( 3,0\right) .  \label{diracop1a}
\end{equation}

The \emph{covariant Dirac equation in STA}, known also as the real Dirac
equation, was obtained by D. Hestenes \cite[VII]{Hes2-03}, \cite[13.3.3,
13.3.3.4]{DorLas}, \cite[6.7, 6.8]{Rodrigues Oliveira} and it is 
\begin{equation}
\hbar \partial \psi \mathrm{I}\mathbf{\sigma }_{3}-\frac{e}{\mathrm{c}}A\psi
=m\mathrm{c}\psi \gamma _{0},\text{ where }\mathrm{I}\mathbf{\sigma }%
_{3}=\gamma _{1}\gamma _{2}.  \label{dirac1}
\end{equation}%
\emph{The real Dirac equation (\ref{dirac1}) is equivalent to the original
Dirac equation}. The equation (\ref{dirac1}) can be recast also as%
\begin{equation}
\left( \mathcal{P}-m\mathrm{c}\overleftarrow{\gamma }_{0}\right) \psi =0,%
\text{ or }\mathcal{P}\psi =m\mathrm{c}\psi \gamma _{0},  \label{dirac1a}
\end{equation}%
where the momentum operator $\mathcal{P}$ and the operator $\overleftarrow{%
\gamma }_{0}$ are defined by%
\begin{equation}
\mathcal{P}\psi =\hbar \partial \psi \mathrm{I}\mathbf{\sigma }_{3}-\frac{e}{%
\mathrm{c}}A\psi ,\qquad \overleftarrow{\gamma }_{0}\psi =\psi \gamma _{0}.
\label{dirac2}
\end{equation}%
The momentum operator $\mathcal{P}$ can be alternatively represented by%
\begin{equation}
\mathcal{P}\psi =\gamma ^{\mu }\mathcal{P}_{\mu }\psi ,\text{ where }%
\mathcal{P}_{\mu }\psi =\hbar \partial _{\mu }\psi \mathrm{I}\mathbf{\sigma }%
_{3}-\frac{e}{\mathrm{c}}A_{\mu }\psi .  \label{dirac2a}
\end{equation}%
We refer to the equations (\ref{dirac1}), (\ref{dirac1a}) as the \emph{%
Dirac-Hestenes equations}. Observe that $\mathcal{P}$ and $\overleftarrow{%
\gamma }_{0}$ commute since the multivectors $\mathrm{I}\mathbf{\sigma }%
_{3}=\gamma _{1}\gamma _{2}$ and $\gamma _{0}$ commute, that is%
\begin{equation}
\left( \mathrm{I}\mathbf{\sigma }_{3}\right) \gamma _{0}=\gamma _{0}\left( 
\mathrm{I}\mathbf{\sigma }_{3}\right) ,\qquad \overleftarrow{\gamma }_{0}%
\mathcal{P}=\mathcal{P}\overleftarrow{\gamma }_{0}.  \label{dirac3}
\end{equation}%
The free electron canonical momentum operator $\mathcal{\mathring{P}}$ is
obtained as a particular case of $\mathcal{P}$ in (\ref{dirac2}) when $A=0$,
that is%
\begin{equation}
\mathcal{\mathring{P}}\psi =\hbar \partial \psi \mathrm{I}\mathbf{\sigma }%
_{3},\text{ and }\mathcal{P}\psi =\mathcal{\mathring{P}}\psi -\frac{e}{%
\mathrm{c}}A\psi =\hbar \partial \psi \mathrm{I}\mathbf{\sigma }_{3}-\frac{e%
}{\mathrm{c}}A\psi .  \label{dirac4}
\end{equation}%
One can also introduce for spinor valued $\psi $ the \emph{covariant
derivative operator} $\mathcal{D}$:%
\begin{equation}
\mathcal{D}\psi =\partial \psi +\frac{e}{\hbar \mathrm{c}}A\psi \mathrm{I}%
\mathbf{\sigma }_{3},\qquad \text{implying }\mathcal{P}=\hbar \partial \psi 
\mathrm{I}\mathbf{\sigma }_{3}-\frac{e}{\mathrm{c}}A\psi \hbar =\mathcal{D}%
\psi \mathrm{I}\mathbf{\sigma }_{3}.  \label{dirac5}
\end{equation}%
Conserved quantities of interest, including the electric current and the
energy-momentum tensor (EnMT), can be obtained from the following \emph{real
Dirac-Hestenes Lagrangian} density for electron in external electromagnetic
field, \cite[4.4]{LasDorGul}, \cite[Ap. B]{Hes-96}, \cite[App. B]{Hes-STC} 
\begin{equation}
L=\mathrm{c}\left\langle \hbar \partial \psi \mathrm{I}\gamma _{3}\tilde{\psi%
}-\frac{e}{\mathrm{c}}A\psi \gamma _{0}\tilde{\psi}-m\mathrm{c}\psi \tilde{%
\psi}\right\rangle .  \label{dirac6}
\end{equation}%
Using expressions (\ref{dirac5}) for the canonical momentum $\mathcal{P}$
and the covariant derivative $\mathcal{D}$, we can transform the
Dirac-Hestenes Lagrangian into the following form%
\begin{equation}
L=\mathrm{c}\left\langle \left[ \left( \mathcal{P}-m\mathrm{c}\overleftarrow{%
\gamma }_{0}\right) \psi \right] \gamma _{0}\tilde{\psi}\right\rangle =%
\mathrm{c}\left\langle \left[ \left( \hbar \mathcal{D}\psi \mathrm{I}\mathbf{%
\sigma }_{3}-m\mathrm{c}\overleftarrow{\gamma }_{0}\right) \psi \right]
\gamma _{0}\tilde{\psi}\right\rangle .  \label{dirac6a}
\end{equation}%
The Lagrangian representation (\ref{dirac6a}) implies%
\begin{equation}
L=0\text{ for any }\psi \text{ satisfying the Dirac equation (\ref{dirac1a}),%
}  \label{dirac6b}
\end{equation}%
and that is typical for the first order systems, \cite[13.3]{DorLas}. One
can also verify that the corresponding Euler-Lagrange field equation is
equivalent to the Dirac-Hestenes equation (\ref{dirac1}).

The \emph{free electron Dirac-Hestenes Lagrangian} $\mathring{L}$ (when $A=0$%
) equals%
\begin{gather}
\mathring{L}=\mathrm{c}\left\langle \hbar \partial \psi \mathrm{I}\gamma _{3}%
\tilde{\psi}-m\mathrm{c}\psi \tilde{\psi}\right\rangle =\mathrm{c}%
\left\langle \left[ \left( \mathcal{\mathring{P}}-m\mathrm{c}\overleftarrow{%
\gamma }_{0}\psi \right) \right] \gamma _{0}\tilde{\psi}\right\rangle =
\label{dirac7} \\
=\mathrm{c}\left\langle \left[ \left( \hbar \partial \psi \mathrm{I}\mathbf{%
\sigma }_{3}-m\mathrm{c}\overleftarrow{\gamma }_{0}\psi \right) \right]
\gamma _{0}\tilde{\psi}\right\rangle ,\text{ where }\mathcal{\mathring{P}}%
_{\mu }\psi =\hbar \partial _{\mu }\psi \mathrm{I}\mathbf{\sigma }_{3}. 
\notag
\end{gather}

\subsection{Conservation laws}

Our treatment of the charge and energy-momentum conservation laws is based
on the Dirac-Hestenes Lagrangian and the Noether theorem.

\subsubsection{Electric charge conservation}

We introduce the so-called \emph{global electromagnetic gauge transformation}
as follows, \cite[3]{Hes-73}, \cite[3.2]{LasDorGul}, 
\begin{equation}
x^{\prime }=x,\qquad \psi ^{\prime }\left( x^{\prime }\right) =\psi \left(
x\right) \mathrm{e}^{\mathrm{I}\mathbf{\sigma }_{3}\epsilon },\quad \epsilon 
\text{ is any real number.}  \label{symch1}
\end{equation}%
Consequently, the global electromagnetic gauge transformation preserves the
vector derivative $\partial \psi $, that is 
\begin{equation}
\partial ^{\prime }=\partial ,\qquad \partial ^{\prime }\psi ^{\prime
}\left( x^{\prime }\right) =\partial \psi \left( x\right) \mathrm{e}^{%
\mathrm{I}\mathbf{\sigma }_{3}\epsilon }.  \label{symch1a}
\end{equation}%
The infinitesimal form of (\ref{symch1}) for for small $\epsilon $ is%
\begin{equation}
\delta x^{\prime }=0,\qquad \bar{\delta}\psi =\psi \mathrm{I}\mathbf{\sigma }%
_{3}\epsilon .  \label{symch2}
\end{equation}%
The \emph{local electromagnetic gauge transformation} is conceived to keep
the covariant derivative $\mathcal{D}\psi $ defined by (\ref{dirac5})
invariant. It involves both the $\psi $ and $A$ and is of the form%
\begin{equation}
x^{\prime }=x,\qquad \psi ^{\prime }\left( x\right) =\psi \left( x\right) 
\mathrm{e}^{\mathrm{I}\mathbf{\sigma }_{3}\frac{e}{\hbar \mathrm{c}}\epsilon
\left( x\right) },\qquad A^{\prime }\left( x\right) =A\left( x\right)
-\partial \epsilon \text{,}  \label{symch3}
\end{equation}%
where $\epsilon \left( x\right) $ is real valued function of $x$ is . Then,
since $\partial =\gamma ^{\mu }\partial _{\mu },$ we consequently obtain 
\begin{equation}
\partial ^{\prime }=\partial ,\qquad \partial ^{\prime }\psi ^{\prime
}\left( x^{\prime }\right) =\left[ \partial \psi \left( x\right) +\frac{e}{%
\hbar \mathrm{c}}\partial \epsilon \psi \left( x\right) \mathrm{I}\mathbf{%
\sigma }_{3}\right] \mathrm{e}^{\mathrm{I}\mathbf{\sigma }_{3}\frac{e}{\hbar 
\mathrm{c}}\epsilon \left( x\right) },  \label{symch4}
\end{equation}%
\begin{equation}
\mathcal{D}^{\prime }\psi ^{\prime }\left( x^{\prime }\right) =\mathcal{D}%
\psi \left( x\right) .  \label{symch5}
\end{equation}%
The infinitesimal form of (\ref{symch3}) for for small $\epsilon \left(
x\right) $ is%
\begin{equation}
\delta x^{\prime }=0,\qquad \bar{\delta}\psi =\psi \mathrm{I}\mathbf{\sigma }%
_{3}\frac{e}{\hbar \mathrm{c}}\epsilon ,\qquad \bar{\delta}\partial \psi
=\partial \psi +\frac{e}{\hbar \mathrm{c}}\partial \epsilon \psi \mathrm{I}%
\mathbf{\sigma }_{3},\qquad \bar{\delta}A=-\partial \epsilon .
\label{symch6}
\end{equation}%
The last equality in (\ref{symch6}) indicates that to have the local gauge
invariance, we have to couple the spinor field $\psi $ with a vector field
to "compensate" for the term $\frac{e}{\hbar \mathrm{c}}\partial \epsilon
\psi \mathrm{I}\mathbf{\sigma }_{3}$. And this exactly what the
electromagnetic potential $A$ does yielding the well known \emph{minimal
coupling}.

One readily verifies that the Dirac-Hestenes Lagrangian (\ref{dirac6}) is
invariant with respect to electromagnetic gauge transformation (\ref{symch1}%
). Then, according to Noether's theorem, there is a conserved electric
current $J^{\mu }$ defined by 
\begin{gather}
\pi ^{\mu }=\frac{\partial L}{\partial \psi _{,\mu }}=\mathrm{c}\hbar 
\mathrm{I}\gamma _{3}\tilde{\psi}\gamma ^{\mu },  \label{ecurr1} \\
J^{\mu }=\pi ^{\mu }\ast \bar{\delta}\psi =\mathrm{c}\left\langle \hbar 
\mathrm{I}\gamma _{3}\tilde{\psi}\gamma ^{\mu }\psi \mathrm{I}\mathbf{\sigma 
}_{3}\epsilon \right\rangle =\mathrm{c}\left\langle \hbar \tilde{\psi}\gamma
^{\mu }\psi \gamma _{0}\epsilon \right\rangle .  \notag
\end{gather}%
Multiplying the current expression in (\ref{ecurr1}) by a proper constant
and using the STA properties (\ref{stag10}), (\ref{scmn1})-(\ref{scmn1b})
together with momentum $\mathcal{P}$ representation (\ref{dirac5}), we
obtain the following expression for the electric current%
\begin{equation}
J^{\mu }=e\mathrm{c}\left\langle \gamma ^{\mu }\psi \gamma _{0}\tilde{\psi}%
\right\rangle ,\qquad J=e\mathrm{c}\psi \gamma _{0}\tilde{\psi},\qquad
J^{\mu }=\left( \mathrm{c}\rho ,\mathbf{J}\right) .  \label{ecurr2}
\end{equation}%
Assuming that $\psi $ satisfies the Dirac equation (\ref{dirac1a}), (\ref%
{dirac2}), we can recast the above expression into%
\begin{equation}
J^{\mu }=\frac{e}{m}\left\langle \gamma ^{\mu }\left( \mathcal{P}\psi
\right) \tilde{\psi}\right\rangle ,\qquad J=J^{\mu }\gamma _{\mu }=\frac{e}{m%
}\left( \mathcal{P}\psi \right) \tilde{\psi}=\frac{e}{m}\left( \chi \partial
\psi \mathrm{I}\mathbf{\sigma }_{3}-\frac{e}{\mathrm{c}}A\psi \right) \tilde{%
\psi}.  \label{ecurr3}
\end{equation}%
The expression $\mathrm{c}\psi \gamma _{0}\tilde{\psi}$ is known as the 
\emph{Dirac probability current} in the QM, \cite[VII.D]{Hes2-03}, \cite[10]%
{Hes-STC}, whereas the expression $\frac{e}{m}\left( \mathcal{P}\psi \right) 
\tilde{\psi}$ is known as the \emph{Gordon current} \cite[5]{Hes-75}, \cite[%
VII.H]{Hes2-03}. \emph{We want to stress that the current }$J$\emph{\
expressions (\ref{ecurr2}) and (\ref{ecurr3}) are evidently two very
different expressions which are equal only because }$\psi $\emph{\ satisfies
the Dirac equation} (\ref{dirac1a}), (\ref{dirac2}). Consequently, \emph{one
may interpret the Dirac equation as a requirement that two generally
different currents defined by (\ref{ecurr2}) and (\ref{ecurr3}) must be the
same.} In addition to that, notice that the expression $\frac{e}{m}\left(
\chi \partial \psi \mathrm{I}\mathbf{\sigma }_{3}-\frac{e}{\mathrm{c}}A\psi
\right) \tilde{\psi}$ in (\ref{ecurr3}) for a general even $\psi $, that is $%
\psi $ not necessarily satisfying the Dirac equation, can take multivector
values. So in the case of general even $\psi $ the proper Gordon current
expression based on its components $J^{\mu }$ is%
\begin{equation}
J=J^{\mu }\gamma _{\mu }=\left\langle \frac{e}{m}\left( \mathcal{P}\psi
\right) \tilde{\psi}\right\rangle _{1}=\frac{e}{m}\left\langle \left( \chi
\partial \psi \mathrm{I}\mathbf{\sigma }_{3}-\frac{e}{\mathrm{c}}A\psi
\right) \tilde{\psi}\right\rangle _{1},  \label{ecurr3a}
\end{equation}%
and in the special case when $\psi $ satisfies the Dirac equation, the
projection operation $\left\langle {}\right\rangle _{1}$ on the vector space
can be naturally omitted since $\left( \chi \partial \psi \mathrm{I}\mathbf{%
\sigma }_{3}-\frac{e}{\mathrm{c}}A\psi \right) \tilde{\psi}$ has to be a
vector in this case. The current $J$ satisfies the conservation law 
\begin{equation}
\partial \cdot J=0\text{ or }\partial _{\mu }J^{\mu }=0,\qquad J^{\mu
}=\left( \mathrm{c}\rho ,\mathbf{J}\right) ,  \label{ecurr3b}
\end{equation}%
where $\rho $ is the charge density and $\mathbf{J}$ is the charge current.

The Gordon current expression (\ref{ecurr3}) satisfies the following \emph{%
Gordon decomposition} law, \cite[3]{Hes-96} 
\begin{equation}
J^{\mu }=J_{\mathrm{c}}^{\mu }+J_{\mathrm{s}}^{\mu },\qquad J_{\mathrm{c}%
}^{\mu }=\frac{e}{m}\left\langle \left( \mathcal{P}^{\mu }\psi \right) 
\tilde{\psi}\right\rangle ,\qquad J_{\mathrm{s}}^{\mu }=\frac{\hbar e}{m}%
\left\langle \left[ \gamma ^{\mu },\gamma ^{\nu }\right] \partial _{\nu
}\psi \mathrm{I}\mathbf{\sigma }_{3}\tilde{\psi}\right\rangle ,
\label{ecurr4}
\end{equation}%
where $J_{\mathrm{c}}^{\mu }$ and $J_{\mathrm{s}}^{\mu }$ \emph{are
respectively the convection and magnetization (spin) currents}. To justify
the use of magnetization and spin terms, let us recall that the \emph{%
magnetization bivector }$M$\emph{\ }and intimately related to it\emph{\ spin
angular momentum bivector }$S$ are defined in the STA by the following
expressions, \cite[4]{Hes-75}, \cite[2, 3]{Hes-96}, \cite[VII.C]{Hes2-03}%
\begin{equation}
S=\frac{\hbar }{2}R\mathrm{I}\mathbf{\sigma }_{3}\tilde{R}=\frac{\hbar }{2}%
R\gamma _{2}\gamma _{1}\tilde{R},\qquad \psi =\varrho ^{\frac{1}{2}}\mathrm{e%
}^{\frac{\mathrm{I}\beta }{2}}R=R\varrho ^{\frac{1}{2}}\mathrm{e}^{\frac{%
\mathrm{I}\beta }{2}},  \label{ecurr5}
\end{equation}%
\begin{equation}
M=\frac{\hbar e}{2m\mathrm{c}}\psi \mathrm{I}\mathbf{\sigma }_{3}\tilde{\psi}%
=\frac{\hbar e}{2m\mathrm{c}}\psi \gamma _{2}\gamma _{1}\tilde{\psi}=\frac{e%
}{m\mathrm{c}}\varrho \mathrm{e}^{\mathrm{I}\beta }S,  \label{ecurr6}
\end{equation}%
where $\psi $ satisfies the canonical relations (\ref{wfudisp1}), (\ref%
{wfudisp2}). Then the following relations between components $J_{\mathrm{s}%
}^{\mu }$ (\ref{ecurr3}) and magnetization bivector $M$ hold:%
\begin{gather}
J_{\mathrm{s}}=J_{\mathrm{s}}^{\mu }\gamma _{\mu }=\mathrm{c}\partial \cdot
M,\qquad J_{\mathrm{s}}^{\mu }=\mathrm{c}\gamma ^{\mu }\cdot \left( \partial
\cdot M\right) =\mathrm{c}\gamma ^{\mu }\cdot \left( \gamma ^{\nu }\cdot
\partial _{\nu }M\right) =  \label{ecurr7} \\
=\mathrm{c}\left( \gamma ^{\mu }\wedge \gamma ^{\nu }\right) \cdot \partial
_{\nu }M=\frac{\hbar e}{m}\left\langle \left( \gamma ^{\mu }\wedge \gamma
^{\nu }\right) \partial _{\nu }\psi \mathrm{I}\mathbf{\sigma }_{3}\tilde{\psi%
}\right\rangle .  \notag
\end{gather}%
It is instructive to see that the STA Gordon current decomposition
representation (\ref{ecurr4})perfectly matches a similar formula in the
conventional Dirac theory, \cite[8.1]{GreinerRQM}, \cite[8.1]{Snygg}, \cite[%
p. 148]{Wachter}:%
\begin{equation}
J^{\mu }=e\mathrm{c}\bar{\Psi}\gamma ^{\mu }\Psi =J_{\mathrm{c}}^{\mu }+J_{%
\mathrm{s}}^{\mu }=\frac{e}{2m}\left[ \bar{\Psi}\hat{P}^{\mu }\Psi -%
\overline{\left( \hat{P}^{\mu }\Psi \right) }\Psi \right] -\frac{\mathrm{i}e%
}{2m}\partial ^{\nu }\left( \bar{\Psi}\sigma _{\ \nu }^{\mu }\Psi \right) ,
\label{ecurr8}
\end{equation}%
where $\sigma _{\ \nu }^{\mu }$ is defined by%
\begin{equation}
\sigma _{\mu \nu }=\frac{\mathrm{i}}{2}\left( \gamma _{\mu }\gamma _{\nu
}-\gamma _{\nu }\gamma _{\mu }\right) ,  \label{ecurr8a}
\end{equation}%
and%
\begin{gather}
J_{\mathrm{c}}^{\mu }=\frac{e}{2m}\left[ \bar{\Psi}\hat{P}^{\mu }\Psi -%
\overline{\left( \hat{P}^{\mu }\Psi \right) }\Psi \right] \text{ is the
convection current density,}  \label{ecurr8b} \\
J_{\mathrm{s}}^{\mu }=-\frac{\mathrm{i}e}{2m}\hat{P}^{\nu }\left( \bar{\Psi}%
\sigma _{\ \nu }^{\mu }\Psi \right) \text{ is the spin current density.} 
\notag
\end{gather}%
Notice that the spin (magnetization) current $J_{\mathrm{s}}$ in view of the
representation $J_{\mathrm{s}}=\partial \cdot M$ in (\ref{ecurr7}) is
conserved since%
\begin{equation}
\partial \cdot J_{\mathrm{s}}=\mathrm{c}\partial \cdot \left( \partial \cdot
M\right) =\mathrm{c}\left( \partial \wedge \partial \right) \cdot M=0.
\label{ecurr8c}
\end{equation}%
Combining (\ref{ecurr3b}) and (\ref{ecurr4}), we obtain also the
conservation law for the convection current $J_{\mathrm{c}}^{\mu }$:%
\begin{equation}
\partial _{\mu }J_{\mathrm{c}}^{\mu }=0.  \label{ecurr8d}
\end{equation}%
Notice also that, assuming that $J$ is the Dirac probability current defined
by (\ref{ecurr2}), we can recast the Lagrangian in (\ref{dirac6a})%
\begin{equation}
L=\mathring{L}-\frac{1}{\mathrm{c}}\left\langle AJ\right\rangle =\mathrm{c}%
\left\langle \left[ \left( \hbar \partial \psi \mathrm{I}\mathbf{\sigma }%
_{3}-m\mathrm{c}\overleftarrow{\gamma }_{0}\psi \right) \right] \gamma _{0}%
\tilde{\psi}\right\rangle -\frac{1}{\mathrm{c}}\left\langle AJ\right\rangle ,
\label{eccurr9}
\end{equation}%
indicating that the current $J$ definition by (\ref{ecurr2}) is in accord
with the classical theory. Indeed, in the classical theory the EM
interaction between the EM field four-potential $A$ and the current $J$ is
described by the expression $\frac{1}{\mathrm{c}}\left\langle
AJ\right\rangle $.

\subsubsection{Energy-momentum conservation}

Applying the Noether theorem to the Dirac Lagrangian (\ref{dirac6}), (\ref%
{dirac6a}) and using relations (\ref{dirac4}), (\ref{dirac6b}), we
consequently obtain for the canonical EnMT $\mathring{T}^{\mu \nu }$ 
\begin{equation}
\pi ^{\mu }=\frac{\partial L}{\partial \psi _{,\mu }}=\mathrm{c}\hbar 
\mathrm{I}\gamma _{3}\tilde{\psi}\gamma ^{\mu },\qquad \mathring{T}^{\mu \nu
}=\pi ^{\mu }\ast \partial ^{\nu }\psi =\mathrm{c}\left\langle \hbar \mathrm{%
I}\gamma _{3}\tilde{\psi}\gamma ^{\mu }\partial ^{\nu }\psi \right\rangle .
\label{dicon1}
\end{equation}%
The above expression for EnMT $\mathring{T}^{\mu \nu }$ after an elementary
transformation turns into%
\begin{equation}
\mathring{T}^{\mu \nu }=\mathrm{c}\left\langle \hbar \mathrm{I}\gamma _{3}%
\tilde{\psi}\gamma ^{\mu }\partial ^{\nu }\psi \right\rangle =\mathrm{c}%
\left\langle \gamma _{0}\tilde{\psi}\gamma ^{\mu }\hbar \partial ^{\nu }\psi 
\mathrm{I}\mathbf{\sigma }_{3}\right\rangle =\mathrm{c}\left\langle \gamma
^{\mu }\left( \mathcal{\mathring{P}}^{\nu }\psi \right) \gamma _{0}\tilde{%
\psi}\right\rangle .  \label{dicon2}
\end{equation}%
Then the EnMT conservation law takes the form 
\begin{equation}
\partial _{\mu }\mathring{T}^{\mu \nu }=-\partial ^{\nu }L=\frac{1}{\mathrm{c%
}}\left( \partial ^{\nu }A^{\mu }\right) J_{\mu }.  \label{dicon3}
\end{equation}%
Observe that the canonical EnMT $\mathring{T}^{\mu \nu }$ involves $\mathcal{%
\mathring{P}}^{\nu }$ and evidently is not gauge invariant. To find its
gauge invariant modification $T^{\mu \nu }$ we use the charge conservation
law $\partial ^{\mu }J_{\mu }=0$ to obtain the following identity%
\begin{gather}
\left( \partial ^{\nu }A^{\mu }\right) J_{\mu }=\left( \partial ^{\nu
}A^{\mu }-\partial ^{\mu }A^{\nu }\right) J_{\mu }+\left( \partial ^{\mu
}A^{\nu }\right) J_{\mu }=  \label{dicon4} \\
=F^{\nu \mu }J_{\mu }+\partial ^{\mu }\left( A^{\nu }J_{\mu }\right) =F^{\nu
\mu }J_{\mu }+\partial _{\mu }\left( A^{\nu }J^{\mu }\right) ,  \notag
\end{gather}%
where $F^{\nu \mu }=\partial ^{\nu }A^{\mu }-\partial ^{\mu }A^{\nu }$ are
components of the EM field bivector $F=\frac{1}{2}F_{\nu \mu }\gamma ^{\nu
}\wedge \gamma ^{\mu }$. The above identity allows to recast the
conservation law (\ref{dicon3}) as%
\begin{equation}
\partial _{\mu }\left( \mathring{T}^{\mu \nu }-\frac{1}{\mathrm{c}}A^{\nu
}J^{\mu }\right) =\frac{1}{\mathrm{c}}F^{\nu \mu }J_{\mu }.  \label{dicon5}
\end{equation}%
The equality (\ref{dicon5}) in turn suggests to introduce the following
gauge invariant modification $T^{\mu \nu }$ of the canonical EnMT $\mathring{%
T}^{\mu \nu }$:%
\begin{equation}
T^{\mu \nu }=\mathring{T}^{\mu \nu }-\frac{1}{\mathrm{c}}A^{\nu }J^{\mu }=%
\mathrm{c}\left\langle \gamma ^{\mu }\left( \mathcal{P}^{\nu }\psi \right)
\gamma _{0}\tilde{\psi}\right\rangle .  \label{dicon6}
\end{equation}%
Then (\ref{dicon5}) can be recast into the conservation law%
\begin{equation}
\partial _{\mu }T^{\mu \nu }=\frac{1}{\mathrm{c}}F^{\nu \mu }J_{\mu },
\label{dicon7}
\end{equation}%
where $\frac{1}{\mathrm{c}}F^{\nu \mu }J_{\mu }$ are the components of the
Lorentz force. Using the identity%
\begin{equation}
F^{\nu \mu }J_{\mu }=\left( \gamma ^{\nu }\wedge \gamma ^{\mu }\right) \cdot
FJ_{\mu }=\gamma ^{\nu }\cdot \left( \gamma ^{\mu }\cdot F\right) J_{\mu
}=\gamma ^{\nu }\cdot \left( J\cdot F\right) ,  \label{dicon8}
\end{equation}%
and introducing the vectors%
\begin{equation}
T^{\mu }=T^{\mu \nu }\gamma _{\nu },  \label{dicon9}
\end{equation}%
we can recast the EnMT conservation (\ref{dicon6}) into a concise vector form%
\begin{equation}
\partial _{\mu }T^{\mu }=\frac{1}{\mathrm{c}}J\cdot F,\text{ where }J\cdot F%
\text{ is the Lorentz force vector.}  \label{dicon10}
\end{equation}%
The properties of the gauge invariant EnMT $T^{\mu \nu }$ and related to $%
T^{\mu }$ are thoroughly studied in \cite[3]{Hes-96}.

\subsection{Free electron solutions to the Dirac equation}

This section provides basic information on the plane wave solutions to the
Dirac-Hestenes equations following to \cite[6]{Hes-81}, \cite[4]{Hes-96}, 
\cite[VIII.B]{Hes2-03}, \cite[8.3.2]{DorLas}. Free electron satisfies the
Dirac equation (\ref{dirac1}) with $A=0$, that is 
\begin{equation}
\hbar \partial \psi \mathrm{I}\mathbf{\sigma }_{3}=m\mathrm{c}\psi \gamma
_{0},\text{ where }\mathrm{I}\mathbf{\sigma }_{3}=\gamma _{1}\gamma _{2}.
\label{frdira1}
\end{equation}%
A \emph{positive energy plane-wave solution} $\psi _{-}$ to the Dirac
equation (\ref{frdira1}) for electron is defined to be of the form%
\begin{equation}
\text{positive energy solution: }\psi _{-}=\psi _{0}\mathrm{e}^{-\mathrm{I}%
\mathbf{\sigma }_{3}k\cdot x},\text{ where }\gamma _{0}\cdot k>0,
\label{frdira2}
\end{equation}%
and $\psi _{0}$ is a constant spinor. Notice that in $\psi _{-}$ the
subindex "$-$" signfies the sign of the electron charge. Recall that the 
\emph{wave vector} $k$ is related to the \emph{momentum vector }$p$ by $%
p=\hbar k$, and we obtain the following spacetime split representations in
terms of relative vectors: 
\begin{equation}
k\gamma _{0}=\frac{\omega }{\mathrm{c}}+\mathbf{k},\qquad p\gamma _{0}=\hbar
k\gamma _{0}=\frac{\hbar \omega }{\mathrm{c}}+\hbar \mathbf{k}=\frac{E}{%
\mathrm{c}}+\mathbf{p}.  \label{frdira2a}
\end{equation}%
If the charge is at rest in the $\gamma _{0}$-frame interpreted as $\mathbf{p%
}=\mathbf{0}$ then according the above formula 
\begin{equation}
p=p\cdot \gamma _{0}=p_{0}=\frac{\hbar \omega _{0}}{\mathrm{c}}=m\mathrm{c.}
\label{fdira2b}
\end{equation}%
Since $\partial =\gamma ^{\mu }\partial _{\mu }$ we have 
\begin{equation}
\partial \psi =\partial \psi _{0}\mathrm{e}^{-\mathrm{I}\mathbf{\sigma }%
_{3}k\cdot x}=-k\psi _{0}\mathrm{e}^{-\mathrm{I}\mathbf{\sigma }_{3}k\cdot x}%
\mathrm{I}\mathbf{\sigma }_{3}=-k\psi \mathrm{I}\mathbf{\sigma }_{3},
\label{frdira3}
\end{equation}%
implying that $\psi =\psi _{0}\mathrm{e}^{-\mathrm{I}\mathbf{\sigma }%
_{3}k\cdot x}$ is a solution to the Dirac equation (\ref{frdira1}) if and
only if $\psi _{0}$ satisfies 
\begin{equation}
p\psi _{0}=m\mathrm{c}\psi _{0}\gamma _{0}.  \label{frdira4}
\end{equation}%
Multiplying the above equation from the right by $\tilde{\psi}_{0}$ we obtain%
\begin{equation}
p\psi _{0}\tilde{\psi}_{0}=m\mathrm{c}\psi _{0}\gamma _{0}\tilde{\psi}_{0}.
\label{frdira5}
\end{equation}%
We assume the constant spinor $\psi _{0}$ to be normalized with the
following canonical representation (\ref{wfudisp2}):%
\begin{gather}
\psi _{0}=\mathrm{e}^{\frac{\mathrm{I}\beta _{0}}{2}}R_{0},\qquad \psi _{0}%
\tilde{\psi}_{0}=\mathrm{e}^{\beta _{0}\mathrm{I}},\text{ where }\beta _{0}%
\text{ is real,}  \label{frdira6} \\
R_{0}\text{ is the Lorentz rotor: }R_{0}\tilde{R}_{0}=\tilde{R}_{0}R_{0}=1. 
\notag
\end{gather}%
Then it follows from (\ref{frdira5}) and (\ref{frdira6}) that%
\begin{equation}
p\mathrm{e}^{\beta _{0}\mathrm{I}}=m\mathrm{c}R_{0}\gamma _{0}\tilde{R}_{0},
\label{frdira7}
\end{equation}%
and since both the $p$ and $R_{0}\gamma _{0}\tilde{R}_{0}$ are vectors, we
must have 
\begin{equation}
\psi _{0}\tilde{\psi}_{0}=\mathrm{e}^{\beta _{0}\mathrm{I}}=\pm 1,\text{
that is }\beta _{0}=0,\pi .  \label{frdira8}
\end{equation}%
Since $\gamma _{0}\cdot p>0$ and $\gamma _{0}\cdot R_{0}\gamma _{0}\tilde{R}%
_{0}>0$ as it follows from (\ref{frdira2}), we must have $\mathrm{e}^{\beta
_{0}\mathrm{I}}=1$ in (\ref{frdira7}), that is%
\begin{equation}
p=m\mathrm{c}R_{0}\gamma _{0}\tilde{R}_{0}.  \label{frdira7a}
\end{equation}%
The rotor $R_{0}$ solving the problem (\ref{frdira7a}) is the product%
\begin{equation}
R_{0}=LU,  \label{frdira7b}
\end{equation}%
where the boost $L$ is defined by%
\begin{equation}
L=\frac{1+v\gamma _{0}}{\left[ 2\left( 1+v\cdot \gamma _{0}\right) \right]
^{1/2}},\qquad v=\frac{p}{m\mathrm{c}}=\gamma \left( 1+\frac{\mathbf{v}}{%
\mathrm{c}}\right) \gamma _{0}=\frac{1}{m\mathrm{c}}\left( \frac{E}{\mathrm{c%
}}+\mathbf{p}\right) \gamma _{0},  \label{frdira7c}
\end{equation}%
or in view of (\ref{frdira2})%
\begin{gather}
L=L\left( \mathbf{p}\right) =\frac{E_{0}+E\left( \mathbf{p}\right) +\mathrm{c%
}\mathbf{p}}{\left[ 2E_{0}\left( E_{0}+E\left( \mathbf{p}\right) \right) %
\right] ^{1/2}},  \label{frdira7d} \\
\text{where }E\left( \mathbf{p}\right) =E_{0}\sqrt{\frac{\mathbf{p}^{2}}{%
\mathrm{c}^{2}}+1},\quad E_{0}=m\mathrm{c}^{2}=p_{0}\mathrm{c}=\hbar k_{0}%
\mathrm{c}=\hbar \omega ,  \label{frdira7e}
\end{gather}%
and the rotor $U$ is a pure rotation in $\gamma _{0}$-frame, that is $%
U\gamma _{0}=\gamma _{0}U$.

A \emph{negative energy plane-wave solution} $\psi _{+}$ to the Dirac
equation (\ref{frdira1}) is defined by a formula similar to (\ref{frdira1})
but with the phase factor $\mathrm{e}^{+\mathrm{I}\mathbf{\sigma }_{3}k\cdot
x}$, namely%
\begin{equation}
\text{negative energy solution: }\psi _{+}=\psi _{0}\mathrm{e}^{\mathrm{I}%
\mathbf{\sigma }_{3}k\cdot x},\text{ where }k\cdot \gamma _{0}>0,
\label{frdira9}
\end{equation}%
Notice that in $\psi _{+}$ the subindex "$+$" signifies that the sign of the
positron charge  is opposite to the negative sign of the electron charge.
For negative energy solutions in place of (\ref{frdira7}) we have%
\begin{equation}
-p\mathrm{e}^{\beta _{0}\mathrm{I}}=m\mathrm{c}R_{0}\gamma _{0}\tilde{R}_{0},
\label{frdira10}
\end{equation}%
and, consequently, $\mathrm{e}^{\beta _{0}\mathrm{I}}=-1$, implying%
\begin{equation}
\text{for negative energy: }p=m\mathrm{c}R_{0}\gamma _{0}\tilde{R}%
_{0},\qquad \psi _{0}\tilde{\psi}_{0}=\mathrm{e}^{\beta _{0}\mathrm{I}}=-1.
\label{frdira11}
\end{equation}%
Positive and negative energy plane wave states are commonly interpreted as
respectively electron state and positron (antiparticle) state with positive
energy, \cite[2.1.6]{Wachter}. Their representations can be summarized by%
\begin{eqnarray}
\text{positive energy (electron)}\text{: } &&\psi _{0}\tilde{\psi}_{0}=%
\mathrm{e}^{\beta _{0}\mathrm{I}}=1\text{:}\qquad \psi _{-}=L\left( \mathbf{p%
}\right) U_{r}\mathrm{e}^{-\mathrm{I}\mathbf{\sigma }_{3}k\cdot x},
\label{frdira12a} \\
\text{negative energy (positron)}\text{: } &&\psi _{0}\tilde{\psi}_{0}=%
\mathrm{e}^{\beta _{0}\mathrm{I}}=-1\text{:}\quad \psi _{+}=L\left( \mathbf{p%
}\right) U_{r}\mathrm{Ie}^{\mathrm{I}\mathbf{\sigma }_{3}k\cdot x},
\label{frdira12b}
\end{eqnarray}%
where $p=\hbar k$ and the subscript $r$ at the spatial rotor $U_{r}$ labels
the spin state with%
\begin{equation}
U_{0}=1,\qquad U_{1}=-\mathrm{I}\mathbf{\sigma }_{2}=\gamma _{1}\gamma
_{3},\qquad U_{1}\gamma _{3}\tilde{U}_{1}=-\gamma _{3}.  \label{frdira13}
\end{equation}

Electron and positron states in (\ref{frdira12a})-(\ref{frdira12b}) can be
related to each other by the so-called \emph{charge conjugation}
transformation, \cite[VII.C, VIII.B]{Hes2-03}, \cite[2.1.6]{Wachter},
defined by%
\begin{equation}
\psi ^{\mathrm{C}}=\psi \mathbf{\sigma }_{2},\text{ where }\mathbf{\sigma }%
_{2}=\gamma _{2}\gamma _{0}.  \label{frdira14}
\end{equation}%
Namely,  $\mathbf{\sigma }_{2}$ anticommutes with $\gamma _{0}$ and $\mathrm{%
I}\mathbf{\sigma }_{3}$, therefore if $\psi $ solves the Dirac equation (\ref%
{dirac1}) with chage $e$ its congjugate $\psi ^{\mathrm{C}}$ solves the
Dirac equation with the charge $-e$, that is%
\begin{equation}
\hbar \partial \psi ^{\mathrm{C}}\mathrm{I}\mathbf{\sigma }_{3}+\frac{e}{%
\mathrm{c}}A\psi ^{\mathrm{C}}=m\mathrm{c}\psi ^{\mathrm{C}}\gamma _{0},%
\text{ where }\mathrm{I}\mathbf{\sigma }_{3}=\gamma _{1}\gamma _{2}.
\label{frdira14a}
\end{equation}%
Notice also that the following identity holds for any real $\alpha $%
\begin{equation}
\mathrm{e}^{-\mathrm{I}\mathbf{\sigma }_{3}\alpha }\mathbf{\sigma }_{2}=%
\mathbf{\sigma }_{2}\mathrm{e}^{\mathrm{I}\mathbf{\sigma }_{3}\alpha },
\label{frdira14b}
\end{equation}%
implying together with (\ref{stag4}), (\ref{sigmg3}) and (\ref{frdira12a})-(%
\ref{frdira12b}) that%
\begin{equation}
\psi _{-}^{\mathrm{C}}=L\left( \mathbf{p}\right) U_{r}^{\prime }\mathrm{Ie}%
^{-\mathrm{I}\mathbf{\sigma }_{3}k\cdot x},\qquad U_{r}^{\prime
}=U_{r}\left( -\mathrm{I}\mathbf{\sigma }_{2}\right) .  \label{frdira14c}
\end{equation}%
Observe that $\psi _{-}^{\mathrm{C}}$ in the above equation is a state
similar to $\psi _{+}$ in (\ref{frdira12b}) indicating that the charge
conjugation transforms an electron state into an antiparticle (positron)
state with positive energy. Note that in view of the last equality in (\ref%
{frdira13}) the factor $-\mathrm{I}\mathbf{\sigma }_{2}=U_{1}$ represents a
spatial rotation that \textquotedblleft flips\textquotedblright\ the
direction of the spin vector, \cite[VIII.B]{Hes2-03}. In fact, the charge
conjugation $\psi \rightarrow \psi ^{\mathrm{C}}$ reverses the charge,
energy, momentum, and spin of an electron state transfering it into a
positron state describing the antiparticle with opposite charge $-e$ in the
same potential $A^{\mu }$, \cite[2.1.6]{Wachter}.

\section{Basics of neoclassical theory of charges with spin of 1/2\label%
{sneobas}}

We develop in this section a spinorial version of our neoclassical field
Lagrangian theory of elementary charges. The initial step in this
development is to assume that the wave function $\psi $ of a single charge
such as electron takes values in the even algebra $\mathrm{Cl}_{+}\left(
1,3\right) $ just as in the Dirac theory. We focus here on the theory of a
single charge in an external electromagnetic field. Extension of this theory
to the case of many elementary charges is similar to the same for spinless
charges constructed and studied in \cite{BF7}-\cite{BF8}.

The Lagrangian of a \emph{single elementary charge in an external
electromagnetic field described by the 4-potential} $\breve{A}$ is 
\begin{equation}
L=\frac{1}{2m}\left\{ \left\langle \mathcal{P}\psi \left( \mathcal{P}\psi
\right) ^{\symbol{126}}\right\rangle -\chi ^{2}\left[ \kappa
_{0}^{2}\left\langle \psi \tilde{\psi}\right\rangle +G\left( \left\langle
\psi \tilde{\psi}\right\rangle \right) \right] \right\} ,\quad \kappa _{0}=%
\frac{m\mathrm{c}}{\chi },  \label{nelag1}
\end{equation}%
where (i) $m$ is the electron mass; (ii) $\chi $ is a constant approximately
equal to the Planck constant $\hbar $; (iii) $G$ \emph{is a nonlinear
self-interaction term of not electromagnetic origin}, and (iv) 
\begin{equation}
\mathcal{P}\psi =\chi \partial \psi \mathrm{I}\mathbf{\sigma }_{3}-\frac{e}{%
\mathrm{c}}\breve{A}\psi  \label{nelag2}
\end{equation}%
is the \emph{momentum operator} which is identical to the same in the
Dirac-Hestenes equation (\ref{dirac1a}). Notice that we have somewhat
departed from the common notations of the Dirac theory denoting the external
EM 4-potential by $\breve{A}$ instead of $A$. The reason for such an
alteration is that \emph{there is no electromagnetic self-interaction for an
elementary charge in our theory}, and every charge is associated with its
individual wave function $\psi $ and elementary EM four potential $A$. So,
to avoid any confusion and to distinguish the external 4-potential from the
elementary 4-potential $A$, we use $\breve{A}$ for the external one.

The nonlinearity $G\left( s\right) $ in (\ref{nelag1}) is defined by the
formula (\ref{nelagG1}). We readily obtain from it%
\begin{equation}
G^{\prime }\left( s\right) =G_{a}^{\prime }\left( s\right) =-a^{-2}\left[
\ln \left( a^{3}\left\vert s\right\vert \right) +\ln \pi ^{3/2}+3\right]
,\quad -\infty <s<\infty .  \label{nelagG2}
\end{equation}%
Notice that (\ref{nelagG1}) and (\ref{nelagG2}) imply the following identity 
\begin{equation}
sG_{a}^{\prime }\left( s\right) -G_{a}\left( s\right) =-a^{-2}s.
\label{nelagG3}
\end{equation}%
As it is already explained the nonlinear self-interaction term $G$ of
non-electromagnetic origin and its role in theory is to provide for the
localization property of the elementary charge in relevant situations.

Just as in the Dirac theory, it is useful to single out the "free" part $%
\mathcal{\mathring{P}}$ of $\mathcal{P}$, namely 
\begin{equation}
\mathcal{P}\psi =\mathcal{\mathring{P}}\psi -\frac{e}{\mathrm{c}}\breve{A}%
\psi \text{, where }\mathcal{\mathring{P}}\psi =\chi \partial \psi \mathrm{I}%
\mathbf{\sigma }_{3}.  \label{nelag2a}
\end{equation}%
The coordinate forms $\mathcal{P}_{\mu }$ and $\mathcal{\mathring{P}}_{\mu }$
of the above momenta operators are%
\begin{gather}
\mathcal{P}_{\mu }\psi =\chi \partial _{\mu }\psi \mathrm{I}\mathbf{\sigma }%
_{3}-\frac{e}{\mathrm{c}}\breve{A}_{\mu }\psi ,\qquad \mathcal{P}\psi
=\left( \gamma ^{\mu }\mathcal{P}_{\mu }\right) \psi ,  \label{nelag2b} \\
\mathcal{\mathring{P}}_{\mu }\psi =\chi \partial _{\mu }\psi \mathrm{I}%
\mathbf{\sigma }_{3},\qquad \mathcal{\mathring{P}}\psi =\left( \gamma ^{\mu }%
\mathcal{\mathring{P}}_{\mu }\right) \psi .  \notag
\end{gather}%
When transforming expressions involving reversion operation, we often use
the following elementary identities: 
\begin{equation}
\widetilde{\mathbf{\sigma }_{3}}=-\mathbf{\sigma }_{3},\qquad \widetilde{%
\mathrm{I}}=\mathrm{I},\qquad \widetilde{\mathrm{I}\mathbf{\sigma }_{3}}=-%
\mathrm{I}\mathbf{\sigma }_{3}=-\mathbf{\sigma }_{3}\mathrm{I}.
\label{psinor3}
\end{equation}%
Lagrangian treatment of the conservation laws based on a multivector
Noether's theorem has been developed in \cite[4-6]{LasDorGul}, \cite[12.4, 13%
]{DorLas}, and we adopt most of that approach here. For more details of
mathematical aspects of the Lagrangian field theory for multivector-valued
fields, we refer the reader to \cite[7]{Rodrigues Oliveira}. To obtain the
Euler-Lagrange equations for the Lagrangian $L$ defined by (\ref{nelag1}),
we find first its derivatives%
\begin{equation}
\frac{\partial L}{\partial \psi }=-\frac{1}{m}\left\{ \left( \mathcal{P}\psi
\right) ^{\symbol{126}}\frac{e}{\mathrm{c}}\breve{A}+\chi ^{2}\left[ \kappa
_{0}^{2}+G^{\prime }\left( \left\langle \psi \tilde{\psi}\right\rangle
\right) \right] \tilde{\psi}\right\} ,  \label{nelag3}
\end{equation}%
\begin{equation}
\pi ^{\mu }=\frac{\partial L}{\partial \psi _{,\mu }}=\frac{\chi }{m}\mathrm{%
I}\mathbf{\sigma }_{3}\left( \mathcal{P}\psi \right) ^{\symbol{126}}\gamma
^{\mu }.  \label{nelag4}
\end{equation}%
Notice that we have dropped the projection operation $\left\langle \ast
\right\rangle _{X}$ in the right-hand sides of (\ref{nelag3}), (\ref{nelag4}%
) since their expressions take values in the even subalgebra. Using
expressions (\ref{nelag3}), (\ref{nelag4}), we obtain the Euler-Lagrange
equation%
\begin{equation}
-\left( \mathcal{P}\psi \right) ^{\symbol{126}}\frac{e}{\mathrm{c}}\breve{A}%
-\chi ^{2}\left[ \kappa _{0}^{2}+G^{\prime }\left( \left\langle \psi \tilde{%
\psi}\right\rangle \right) \right] \tilde{\psi}-\partial _{\mu }\chi \mathrm{%
I}\mathbf{\sigma }_{3}\left( \mathcal{P}\psi \right) ^{\symbol{126}}\gamma
^{\mu }=0.  \label{nelag5}
\end{equation}%
Application of the reversion operation to the above equation yields%
\begin{equation}
-\frac{e}{\mathrm{c}}\breve{A}\mathcal{P}\psi -\left[ \kappa
_{0}^{2}+G^{\prime }\left( \left\langle \psi \tilde{\psi}\right\rangle
\right) \right] \psi +\chi \partial \left( \mathcal{P}\psi \right) \mathrm{I}%
\mathbf{\sigma }_{3}=0,  \label{nelag5a}
\end{equation}%
which, in turn, in view of the expression (\ref{nelag2}) for $\mathcal{P}$,
can be transformed into a more concise form of the \emph{field equation}%
\begin{equation}
\mathcal{P}^{2}\psi -\left[ \kappa _{0}^{2}+G^{\prime }\left( \left\langle
\psi \tilde{\psi}\right\rangle \right) \right] \psi =0.  \label{nelag6}
\end{equation}%
\emph{Hence, the field equation (\ref{nelag6}) is the master evolution
equation for the wave function in our theory based on the Lagrangian (\ref%
{nelag1}).} The expression $\mathcal{P}^{2}\psi $ in equation (\ref{nelag6})
can be transformed into the following form showing the external EM field%
\begin{equation}
\mathcal{P}^{2}\psi =\mathcal{\mathring{P}}^{2}\psi -\frac{\chi e}{c}\left[
F+2\breve{A}\cdot \partial \right] \psi \mathrm{I}\mathbf{\sigma }_{3}+\frac{%
e^{2}}{c^{2}}\breve{A}^{2}\psi ,  \label{nelag6e}
\end{equation}%
where $F=\partial \wedge \breve{A}$ is the bivector of the electromagnetic
field.

Using the commutativity (\ref{dirac3}) of the operator $\overleftarrow{%
\gamma }_{0}$ and the momentum operator $\mathcal{P}$, one can factorize the
expression $\mathcal{P}^{2}\psi -\kappa _{0}^{2}\psi $ in the equation (\ref%
{nelag6}) yielding \emph{\ }%
\begin{equation}
\left( \mathcal{P}+m\mathrm{c}\overleftarrow{\gamma }_{0}\right) \left( 
\mathcal{P}-m\mathrm{c}\overleftarrow{\gamma }_{0}\right) \psi -\chi
^{2}G^{\prime }\left( \left\langle \psi \tilde{\psi}\right\rangle \right)
\psi =0.  \label{nelag6f}
\end{equation}%
It is instructive to compare the\ above field equation (\ref{nelag6}) with
the Dirac-Hestenes equation (\ref{dirac1a}), (\ref{dirac2}). Just by looking
at the two equations, one can see two significant differences. First of all,
the field equation (\ref{nelag6}) contains a nonlinear self-interaction term 
$G^{\prime }\left( \left\langle \psi \tilde{\psi}\right\rangle \right) $,
that is a concept not present in the Dirac theory. For comparison purposes
it is instructive to eliminate this nonlinear term from the field equation (%
\ref{nelag6}) resulting in%
\begin{equation}
\text{the truncated field equation: }\left( \mathcal{P}^{2}-m^{2}\mathrm{c}%
^{2}\right) \psi =0.  \label{nelag6a}
\end{equation}%
Now one can see another significant difference between the truncated field
equation (\ref{nelag6a}) and the Dirac-Hestenes equation (\ref{dirac1a}).
Indeed, the Dirac-Hestenes equation (\ref{dirac1a}) is linear in $\mathcal{P}
$ whereas the truncated field equation (\ref{nelag6a}) is quadratic in $%
\mathcal{P}$. In spite of this difference it is possible to establish an
intimate relation between the two equations by factorizing the truncated
field equation (\ref{nelag6a}). To do that we use the commutativity (\ref%
{dirac3}) of the operator $\overleftarrow{\gamma }_{0}$ and the momentum
operator $\mathcal{P}$ and factorize equation (\ref{nelag6}) into the
following form%
\begin{equation}
\left( \mathcal{P}+m\mathrm{c}\overleftarrow{\gamma }_{0}\right) \left( 
\mathcal{P}-m\mathrm{c}\overleftarrow{\gamma }_{0}\right) \psi =0,\text{
truncated field equation factorized.}  \label{nelag7}
\end{equation}%
The above factorization of the truncated field equation is not unique. In
fact, one can drop the operator $\overleftarrow{\gamma }_{0}$ from it, and
what is left is still a correct representation of the original field
equation (\ref{nelag6}). An important justification for the factorization (%
\ref{nelag7}) with the operator $\overleftarrow{\gamma }_{0}$ is as follows.
For even $\psi $ both the vectors $\mathcal{P}\psi $ and $m\mathrm{c}%
\overleftarrow{\gamma }_{0}\psi $ are odd and hence each of the equation%
\begin{equation}
\left( \mathcal{P}-m\mathrm{c}\overleftarrow{\gamma }_{0}\right) \psi
=0,\qquad \left( \mathcal{P}+m\mathrm{c}\overleftarrow{\gamma }_{0}\right)
\psi =0  \label{nelag8}
\end{equation}%
can have even solutions. On the other hand, the equations%
\begin{equation}
\left( \mathcal{P}-m\mathrm{c}\right) \psi =0,\qquad \left( \mathcal{P}+m%
\mathrm{c}\right) \psi =0  \label{nelag9}
\end{equation}%
can not have a nontrivial even solution since for even $\psi $ the
multivector $\mathcal{P}\psi $ is always odd.

Observe now that any linear combination of solutions to equations (\ref%
{nelag8}) is a solution to the truncated field equations (\ref{nelag7}).
Hence any solution to the Dirac-Hestenes equation (\ref{dirac1})-(\ref%
{dirac2}) solves also the truncated form (\ref{nelag7}) of the neoclassical
field equation. In particular, let us take the external potential $\hat{A}$
to be the Coulomb potential, that is $\hat{A}=\hat{A}_{\mathrm{c}}^{\mu
}=\left( -\frac{Ze^{2}}{\left\vert \mathbf{x}\right\vert },0\right) $ where $%
Z$ is the nucleus charge. Then solutions to the Dirac-Hestenes equation for
the Coulomb potential%
\begin{equation}
\hbar \partial \psi \mathrm{I}\mathbf{\sigma }_{3}-\frac{e}{\mathrm{c}}\hat{A%
}_{\mathrm{c}}\psi =m\mathrm{c}\psi \gamma _{0},\qquad \hat{A}_{\mathrm{c}%
}^{\mu }=\left( -\frac{Ze^{2}}{\left\vert \mathbf{x}\right\vert },0\right) ,
\label{nelag9a}
\end{equation}%
are solutions to the truncated field equation (\ref{nelag7}) and,
consequently, are approximate solutions to the neoclassical field equation (%
\ref{nelag6}) with neglected nonlinearity $G$. Notice that the typical
spatial scale of electron states in the Coulomb potential is the Bohr
radius, and if the size parameter $a$ is much larger than the Bohr radius,
then the nonlinearity can be neglected, see \cite{BF7}. Since the
Dirac-Hestenes equation for the Coulomb potential (\ref{nelag9a}) is exactly
equivalent to the original Dirac equation for the same potential, \cite[VII]%
{Hes2-03}, we can claim the that frequency spectrum of the neoclassical
field equation (\ref{nelag6}) includes as an approximation the well known
frequency spectrum of the Dirac equation, \cite[8.2]{Schwabl}.

Interestingly, in \cite{SanMar} the equation (\ref{nelag6a}) (called the
"square of the Dirac equation") is derived by conformal differential
geometry. The general setup in \cite{SanMar} , though very different from
our neoclassical approach, has some common features including the underlying
continuum and that the QM is not a starting point but rather an
approximation.

\subsection{Conservation laws}

Our treatment of the charge and energy-momentum conservation is based on the
Noether theorem and consequently requires the knowledge of relevant groups
of transformations which leave the Lagrangian invariant.

\subsubsection{Charge and current densities}

The neoclassical Lagrangian (\ref{nelag3}) is invariant with respect to the
global charge gauge transformation (\ref{symch2}) as in the case of the
Dirac theory. Consequently, Noether's current reduces in this case to the
following expression for the electric current%
\begin{gather}
J^{\mu }=\pi ^{\mu }\ast \bar{\delta}\psi =\frac{\partial \mathcal{L}}{%
\partial \psi _{,\mu }}\ast \bar{\delta}\psi =\frac{\chi }{m}\left\langle 
\mathrm{I}\mathbf{\sigma }_{3}\left( \mathcal{P}\psi \right) ^{\symbol{126}%
}\gamma ^{\mu }\psi \mathrm{I}\mathbf{\sigma }_{3}\epsilon \right\rangle =
\label{Jpiem1} \\
=-\frac{\chi }{m}\left\langle \left( \mathcal{P}\psi \right) ^{\symbol{126}%
}\gamma ^{\mu }\psi \right\rangle \epsilon =-\frac{\chi }{m}\left\langle
\gamma ^{\mu }\left( \mathcal{P}\psi \right) \tilde{\psi}\right\rangle
\epsilon ,  \notag
\end{gather}%
where we have used expressions (\ref{nelag4}) and (\ref{symch2}) for $\pi
^{\mu }$ and $\bar{\delta}\psi $ respectively. Multiplication of the above
expression for $J^{\mu }$ by a suitable constant and consequent
transformations yield the following expressions for the current components 
\begin{gather}
J^{\mu }=\frac{e}{m}\left\langle \gamma ^{\mu }\left( \mathcal{P}\psi
\right) \tilde{\psi}\right\rangle =\frac{e}{m}\gamma ^{\mu }\cdot
\left\langle \left( \mathcal{P}\psi \right) \tilde{\psi}\right\rangle _{1}=
\label{Jpiem2} \\
=\frac{e}{m}\left\langle \gamma ^{\mu }\left( \chi \partial \psi \mathrm{I}%
\mathbf{\sigma }_{3}-\frac{e}{\mathrm{c}}\breve{A}\psi \right) \tilde{\psi}%
\right\rangle =\frac{e}{m}\gamma ^{\mu }\cdot \left\langle \left( \chi
\partial \psi \mathrm{I}\mathbf{\sigma }_{3}-\frac{e}{\mathrm{c}}\breve{A}%
\psi \right) \tilde{\psi}\right\rangle _{1},  \notag
\end{gather}%
implying the following concise form for the current vector%
\begin{equation}
J=\frac{e}{m}\gamma _{\mu }J^{\mu }=\frac{e}{m}\left\langle \left( \mathcal{P%
}\psi \right) \tilde{\psi}\right\rangle _{1}=\frac{e}{m}\left\langle \left(
\chi \partial \psi \mathrm{I}\mathbf{\sigma }_{3}-\frac{e}{\mathrm{c}}\breve{%
A}\psi \right) \tilde{\psi}\right\rangle _{1}.  \label{Jpiem3}
\end{equation}%
\emph{Observe that our expressions (\ref{Jpiem2}) and (\ref{Jpiem3}) for the
current }$J$\emph{\ are exactly the same as the current expressions (\ref%
{ecurr3}) and (\ref{ecurr3a}) in the Dirac theory}

Notice that any even $\psi $ according to (\ref{wfudisp1}) satisfies $\psi 
\tilde{\psi}=\tilde{\psi}\psi =\varrho \mathrm{e}^{\beta \mathrm{I}}$.
Consequently, for any vector $\breve{A}$ we have 
\begin{gather}
\left\langle \tilde{\psi}\gamma ^{\mu }\breve{A}\psi \right\rangle
=\left\langle \gamma ^{\mu }\breve{A}\psi \tilde{\psi}\right\rangle
=\left\langle \left( \gamma ^{\mu }\cdot \breve{A}+\gamma ^{\mu }\wedge 
\breve{A}\right) \rho \mathrm{e}^{\beta \mathrm{I}}\right\rangle =
\label{Jpiem3a} \\
=\left\langle \gamma ^{\mu }\cdot \breve{A}\rho \mathrm{e}^{\beta \mathrm{I}%
}\right\rangle =\breve{A}^{\mu }\left\langle \rho \mathrm{e}^{\beta \mathrm{I%
}}\right\rangle =\breve{A}^{\mu }\left\langle \tilde{\psi}\psi \right\rangle
.  \notag
\end{gather}%
Using the above identity we can transform the current components $J^{\mu }$
in (\ref{Jpiem2}) as follows: 
\begin{gather}
\left\langle \gamma ^{\mu }\left( \mathcal{P}\psi \right) \tilde{\psi}%
\right\rangle =\left\langle \gamma ^{\mu }\left( \chi \partial \psi \mathrm{I%
}\mathbf{\sigma }_{3}-\frac{e}{\mathrm{c}}\breve{A}\psi \right) \tilde{\psi}%
\right\rangle =\chi \left\langle \gamma ^{\mu }\gamma ^{\nu }\partial _{\nu
}\psi \mathrm{I}\mathbf{\sigma }_{3}\tilde{\psi}\right\rangle -\left\langle 
\frac{e}{\mathrm{c}}\breve{A}^{\mu }\psi \tilde{\psi}\right\rangle =
\label{Jpiem4} \\
=\frac{\chi }{2}\left\langle \left( \gamma ^{\mu }\gamma ^{\nu }+\gamma
^{\nu }\gamma ^{\mu }\right) \partial _{\nu }\psi \mathrm{I}\mathbf{\sigma }%
_{3}\tilde{\psi}\right\rangle +\frac{\chi }{2}\left\langle \left( \gamma
^{\mu }\gamma ^{\nu }-\gamma ^{\nu }\gamma ^{\mu }\right) \partial _{\nu
}\psi \mathrm{I}\mathbf{\sigma }_{3}\tilde{\psi}\right\rangle -\left\langle 
\tilde{\psi}\frac{e}{\mathrm{c}}\breve{A}^{\mu }\psi \right\rangle =  \notag
\\
=\frac{\chi }{2}\left\langle 2g^{\mu \nu }\partial _{\nu }\psi \mathrm{I}%
\mathbf{\sigma }_{3}\tilde{\psi}\right\rangle +\chi \left\langle \left(
\gamma ^{\mu }\wedge \gamma ^{\nu }\right) \partial _{\nu }\psi \mathrm{I}%
\mathbf{\sigma }_{3}\tilde{\psi}\right\rangle -\left\langle \frac{e}{\mathrm{%
c}}\breve{A}^{\mu }\psi \tilde{\psi}\right\rangle =  \notag \\
=\left\langle \left( \mathcal{P}^{\mu }\psi \right) \tilde{\psi}%
\right\rangle +\chi \left\langle \left( \gamma ^{\mu }\wedge \gamma ^{\nu
}\right) \partial _{\nu }\psi \mathrm{I}\mathbf{\sigma }_{3}\tilde{\psi}%
\right\rangle ,  \notag
\end{gather}%
yielding%
\begin{equation}
J^{\mu }=\frac{e}{m}\left\langle \left( \mathcal{P}^{\mu }\psi \right) 
\tilde{\psi}\right\rangle +\frac{\chi e}{m}\left\langle \left[ \gamma ^{\mu
},\gamma ^{\nu }\right] \partial _{\nu }\psi \mathrm{I}\mathbf{\sigma }_{3}%
\tilde{\psi}\right\rangle ,\text{ where }\mathcal{P}^{\mu }\psi =\chi
\partial ^{\mu }\psi \mathrm{I}\mathbf{\sigma }_{3}-\frac{e}{\mathrm{c}}%
\breve{A}^{\mu }\psi ,  \label{Jpiem5}
\end{equation}%
where we used the commutator product $\left[ \gamma ^{\mu },\gamma ^{\nu }%
\right] $ notation. Observe that the current expression (\ref{Jpiem5}) is
exactly the same as the Gordon current decomposition (\ref{ecurr4}) for the
current in the Dirac theory if we substitute $\hbar $ with $\chi $, namely 
\begin{equation}
J^{\mu }=J_{\mathrm{c}}^{\mu }+J_{\mathrm{s}}^{\mu },\qquad J_{\mathrm{c}%
}^{\mu }=\frac{e}{m}\left\langle \left( \mathcal{P}^{\mu }\psi \right) 
\tilde{\psi}\right\rangle ,\qquad J_{\mathrm{s}}^{\mu }=\frac{\chi e}{m}%
\left\langle \left[ \gamma ^{\mu },\gamma ^{\nu }\right] \partial _{\nu
}\psi \mathrm{I}\mathbf{\sigma }_{3}\tilde{\psi}\right\rangle ,
\label{Jpiem6}
\end{equation}%
where $J_{\mathrm{c}}^{\mu }$ and $J_{\mathrm{s}}^{\mu }$ \emph{are
respectively the convection and magnetization (spin) currents}.
Consequently, just as in the case of the Dirac theory as indicated by
conservation laws (\ref{ecurr8c}), (\ref{ecurr8d}), these currents are
conserved individually%
\begin{equation}
\partial _{\mu }J_{\mathrm{c}}^{\mu }=0,\qquad \partial \cdot J_{\mathrm{s}%
}=0.  \label{Jpiem7}
\end{equation}%
Notice that the representation (\ref{ecurr5})-(\ref{ecurr7}) for the
magnetization/spin current in the Dirac theory holds in the neoclassical
case as well, namely%
\begin{gather}
J_{\mathrm{s}}=J_{\mathrm{s}}^{\mu }\gamma _{\mu }=\mathrm{c}\partial \cdot
M,\qquad J_{\mathrm{s}}^{\mu }=\mathrm{c}\gamma ^{\mu }\cdot \left( \partial
\cdot M\right) =\mathrm{c}\gamma ^{\mu }\cdot \left( \gamma ^{\nu }\cdot
\partial _{\nu }M\right) =  \label{Jpiem8} \\
=\mathrm{c}\left( \gamma ^{\mu }\wedge \gamma ^{\nu }\right) \cdot \partial
_{\nu }M=\frac{\hbar e}{m}\left\langle \left( \gamma ^{\mu }\wedge \gamma
^{\nu }\right) \partial _{\nu }\psi \mathrm{I}\mathbf{\sigma }_{3}\tilde{\psi%
}\right\rangle .  \notag
\end{gather}%
Observe also that the \emph{magnetization bivector }$M$ defined in (\ref%
{KGfreg6})\emph{\ }can be related to the \emph{spin angular momentum
bivector }$S$ as follows, \cite[4]{Hes-75}, \cite[2, 3]{Hes-96}, \cite[VII.C]%
{Hes2-03} 
\begin{equation}
S=\frac{\hbar }{2}R\mathrm{I}\mathbf{\sigma }_{3}\tilde{R}=\frac{\hbar }{2}%
R\gamma _{2}\gamma _{1}\tilde{R},\qquad \psi =\varrho ^{\frac{1}{2}}\mathrm{e%
}^{\frac{\mathrm{I}\beta }{2}}R=R\varrho ^{\frac{1}{2}}\mathrm{e}^{\frac{%
\mathrm{I}\beta }{2}},  \label{Jpiem9}
\end{equation}%
\begin{equation}
M=\frac{\hbar e}{2m\mathrm{c}}\psi \mathrm{I}\mathbf{\sigma }_{3}\tilde{\psi}%
=\frac{\hbar e}{2m\mathrm{c}}\psi \gamma _{2}\gamma _{1}\tilde{\psi}=\frac{e%
}{m\mathrm{c}}\mathrm{e}^{\mathrm{I}\beta }\varrho S.  \label{Jpiem10}
\end{equation}

\subsubsection{Gauge invariant energy-momentum tensor}

In the case of the neoclassical Lagrangian (\ref{nelag3}), the general
expression for the canonical EnMT $\mathring{T}^{\mu \nu }$ with the help of
(\ref{nelag4}) reduces to 
\begin{equation}
\mathring{T}^{\mu \nu }=\pi ^{\mu }\ast \partial ^{\nu }\psi -\delta ^{\mu
\nu }L,\quad \text{where }\pi ^{\mu }=\frac{\partial L}{\partial \psi _{,\mu
}}=\frac{\chi }{m}\mathrm{I}\mathbf{\sigma }_{3}\left( \mathcal{P}\psi
\right) ^{\symbol{126}}\gamma ^{\mu },  \label{Tmunp1}
\end{equation}%
implying%
\begin{equation}
\mathring{T}^{\mu \nu }=\frac{\chi }{m}\left\langle \gamma ^{\mu }\partial
^{\nu }\psi \mathrm{I}\mathbf{\sigma }_{3}\left( \mathcal{P}\psi \right) ^{%
\symbol{126}}\right\rangle -\delta _{\nu }^{\mu }L,\quad \text{where }%
\mathcal{P}\psi =\chi \partial \psi \mathrm{I}\mathbf{\sigma }_{3}-\frac{e}{%
\mathrm{c}}\breve{A}\psi .  \label{Tmunp2}
\end{equation}%
The corresponding conservation law takes then the form%
\begin{equation}
\partial _{\mu }\mathring{T}^{\mu \nu }=-\partial ^{\nu }L,\text{ where }%
\partial ^{\nu }L=\frac{\partial L}{\partial x_{\nu }}.  \label{Tmunp3}
\end{equation}%
Since the explicit dependence on $x_{\nu }$ in $L$ comes only through the EM
potential $\breve{A},$ we find that%
\begin{eqnarray}
\partial ^{\nu }L &=&\partial ^{\nu }\frac{1}{2m}\left\langle \left( 
\mathcal{P}\psi \right) \left( \mathcal{P}\psi \right) ^{\symbol{126}%
}\right\rangle =\frac{1}{m}\left\langle \left[ \partial ^{\nu }\left( 
\mathcal{P}\psi \right) \right] \left( \mathcal{P}\psi \right) ^{\symbol{126}%
}\right\rangle =  \label{Tmunp4} \\
&=&-\frac{1}{m}\frac{e}{\mathrm{c}}\left\langle \left( \partial ^{\nu }%
\breve{A}\right) \psi \left( \mathcal{P}\psi \right) ^{\symbol{126}%
}\right\rangle =-\frac{1}{m}\frac{e}{\mathrm{c}}\left( \partial ^{\nu }%
\breve{A}^{\mu }\right) \left\langle \gamma _{\mu }\psi \left( \mathcal{P}%
\psi \right) ^{\symbol{126}}\right\rangle =  \notag \\
&=&-\frac{1}{\mathrm{c}}\frac{e}{m}\left( \partial ^{\nu }\breve{A}^{\mu
}\right) \left\langle \gamma _{\mu }\left( \mathcal{P}\psi \right) \tilde{%
\psi}\right\rangle =-\frac{1}{\mathrm{c}}\left( \partial ^{\nu }\breve{A}%
^{\mu }\right) J_{\mu }.  \notag
\end{eqnarray}%
The canonical EnMT $\mathring{T}^{\mu \nu }$ defined by (\ref{Tmunp2}) is
evidently not gauge invariant. To modify it into a gauge invariant form, we
use the expression (\ref{Jpiem2}) for the current components $J^{\mu }$ and
transform EnMT $\mathring{T}^{\mu \nu }$ as follows:%
\begin{gather}
\mathring{T}^{\mu \nu }+\delta _{\nu }^{\mu }L=\frac{\chi }{m}\left\langle
\gamma ^{\mu }\partial ^{\nu }\psi \mathrm{I}\mathbf{\sigma }_{3}\left( 
\mathcal{P}\psi \right) ^{\symbol{126}}\right\rangle =\frac{1}{m}%
\left\langle \gamma ^{\mu }\chi \partial ^{\nu }\psi \mathrm{I}\mathbf{%
\sigma }_{3}\left( \mathcal{P}\psi \right) ^{\symbol{126}}\right\rangle =
\label{Tmunp5} \\
=\frac{1}{m}\left\langle \gamma ^{\mu }\left( \mathcal{P}^{\nu }\psi \right)
\left( \mathcal{P}\psi \right) ^{\symbol{126}}\right\rangle +\frac{1}{m}%
\left\langle \gamma ^{\mu }\frac{e}{\mathrm{c}}\breve{A}^{\nu }\psi \left( 
\mathcal{P}\psi \right) ^{\symbol{126}}\right\rangle =  \notag \\
=\frac{1}{m}\left\langle \gamma ^{\mu }\left( \mathcal{P}^{\nu }\psi \right)
\left( \mathcal{P}\psi \right) ^{\symbol{126}}\right\rangle +\frac{1}{%
\mathrm{c}}\breve{A}^{\nu }\frac{e}{m}\left\langle \gamma ^{\mu }\left( 
\mathcal{P}\psi \right) \tilde{\psi}\right\rangle =  \notag \\
=\frac{1}{m}\left\langle \gamma ^{\mu }\left( \mathcal{P}^{\nu }\psi \right)
\left( \mathcal{P}\psi \right) ^{\symbol{126}}\right\rangle +\frac{1}{%
\mathrm{c}}\breve{A}^{\nu }J^{\mu }.  \notag
\end{gather}%
The above equality suggests to introduce the following expression for a
gauge invariant EnMT $T^{\mu \nu }$:%
\begin{equation}
T^{\mu \nu }=\mathring{T}^{\mu \nu }-\frac{1}{\mathrm{c}}\breve{A}^{\nu
}J^{\mu }=\frac{1}{m}\left\langle \gamma ^{\mu }\left( \mathcal{P}^{\nu
}\psi \right) \left( \mathcal{P}\psi \right) ^{\symbol{126}}\right\rangle
-\delta _{\nu }^{\mu }L.  \label{Tmunp6}
\end{equation}%
Indeed, using the conservation law $\partial _{\mu }J^{\mu }=0$ and the
canonical EnMT $\mathring{T}^{\mu \nu }$ conservation law (\ref{Tmunp3}), we
obtain%
\begin{eqnarray}
\partial _{\mu }T^{\mu \nu } &=&\partial _{\mu }\mathring{T}^{\mu \nu }-%
\frac{1}{\mathrm{c}}\left( \partial _{\mu }\breve{A}^{\nu }\right) J^{\mu
}=-\partial ^{\nu }L-\frac{1}{\mathrm{c}}\left( \partial _{\mu }\breve{A}%
^{\nu }\right) J^{\mu }=  \label{Tmunp7} \\
&=&\frac{1}{\mathrm{c}}\left( \partial ^{\nu }\breve{A}^{\mu }\right) J_{\mu
}-\frac{1}{\mathrm{c}}\left( \partial ^{\mu }\breve{A}^{\nu }\right) J_{\mu
}=\frac{1}{\mathrm{c}}F^{\nu \mu }J_{\mu },  \notag
\end{eqnarray}%
where $\frac{1}{\mathrm{c}}F^{\nu \mu }J_{\mu }$ is the Lorentz force.

\section{Neoclassical free charge with spin\label{sneofree}}

In this section we carry out a rather detailed analysis of the basic case of
the free charge when $\breve{A}=0$. The charge Lagrangian (\ref{nelag1}) in
the case of the free charge takes the form%
\begin{equation}
L=\frac{\chi ^{2}}{2m}\left\{ \left\langle \gamma ^{\beta }\gamma ^{\alpha
}\left( \partial _{\alpha }\psi \right) \left( \partial _{\beta }\tilde{\psi}%
\right) \right\rangle -\left[ \kappa _{0}^{2}\left\langle \psi \tilde{\psi}%
\right\rangle +G\left( \left\langle \psi \tilde{\psi}\right\rangle \right) %
\right] \right\} ,\quad \kappa _{0}=\frac{m\mathrm{c}}{\chi }.
\label{KGfreg0}
\end{equation}%
The field equation (\ref{nelag6}) when $\breve{A}=0$ after straightforward
transformations turns into \emph{free charge spinor field equation}%
\begin{equation}
-\chi ^{2}\partial _{\mu }\partial ^{\mu }\psi -\left[ \kappa
_{0}^{2}+G^{\prime }\left( \left\langle \psi \tilde{\psi}\right\rangle
\right) \right] \psi =0.  \label{KGfreg1}
\end{equation}%
The above equation is similar to a \emph{scalar nonlinear Klein-Gordon}
(NKG) equation that arises in our neoclassical scalar theory, \cite{BF8},
namely 
\begin{equation}
-\frac{1}{\mathrm{c}^{2}}\partial _{t}^{2}\psi +\nabla ^{2}\psi -\kappa
_{0}^{2}\psi -G^{\prime }\left( \left\vert \psi \right\vert ^{2}\right) \psi
=0,  \label{KGfreg2}
\end{equation}%
where $\psi $ is complex-valued. The scalar equation (\ref{KGfreg2}) and its
solutions are relevant to the analysis of the spinor field equation (\ref%
{KGfreg1}) and its basic properties are considered in the following section.

\subsection{Scalar equation}

Equation (\ref{KGfreg2}) is the Euler-Lagrange equation associated with the
Lagrangian 
\begin{equation}
L=\frac{\chi ^{2}}{2m}\left\{ \partial _{\mu }\psi ^{\ast }\partial ^{\mu
}\psi -\left[ \kappa _{0}^{2}\psi ^{\ast }\psi +G\left( \psi ^{\ast }\psi
\right) \right] \right\} ,  \label{KGfreg3}
\end{equation}%
where the nonlinearity $G=G_{a}$ is defined be equation (\ref{nelagG1}) for $%
s\geq 0$. The fundamental rest solution to the scalar NKG equation (\ref%
{KGfreg2}) is of the form, \cite{BF8} 
\begin{gather}
\psi _{\pm }\left( t,\mathbf{x}\right) =\mathrm{e}^{\mp \mathrm{i}\omega
_{0}t}\mathring{u}\left( \left\vert \mathbf{x}\right\vert \right) ,\text{
where }\omega _{0}=\frac{m\mathrm{c}^{2}}{\chi }=\kappa _{0}\mathrm{c},
\label{KGfreg4} \\
\mathring{u}\left( s\right) =\mathring{u}_{a}\left( s\right) =a^{-3/2}\pi
^{-3/4}\exp \left( -\frac{s^{2}}{2a^{2}}\right) ,\qquad s\geq 0,
\label{KGfreg4a}
\end{gather}%
where $\mathring{u}\left( \left\vert \mathbf{x}\right\vert \right) $
satisfies the equation 
\begin{equation}
\nabla ^{2}\mathring{u}\left( \left\vert \mathbf{x}\right\vert \right)
-G^{\prime }\left( \mathring{u}^{2}\left( \left\vert \mathbf{x}\right\vert
\right) \right) \mathring{u}\left( \left\vert \mathbf{x}\right\vert \right)
=0.  \label{KGfreg4c}
\end{equation}%
Observe that the solution $\psi _{\pm }$ is the product of a time-harmonic
factor $\mathrm{e}^{\mp \mathrm{i}\omega _{0}t}$ and the Gaussian factor $%
\mathring{u}\left( \left\vert \mathbf{x}\right\vert \right) $ describing the
localized shape of the wave. An STA representation of the above solution
which is manifestly coordinate free is as follows: 
\begin{equation}
\psi \left( x\right) =\psi _{\mp }\left( v,x\right) =\mathrm{e}^{\mp \mathrm{%
i}\kappa _{0}x\cdot v}\mathring{u}\left( \sqrt{\left( x\cdot v\right)
^{2}-x^{2}}\right) ,\qquad \kappa _{0}=\frac{\omega _{0}}{\mathrm{c}},
\label{KGfreg4b}
\end{equation}%
where $v$ is proper velocity of the free electron, and one can think of $v$
as describing the rest frame of the electron.

It is instructive to find a representation of the solution $\psi \left(
v,x\right) $ in (\ref{KGfreg4b}) in the frame of an arbitrary inertial
observer $\gamma _{0}$ by relating it to the inertial observer $v=\gamma
_{0}^{\prime }$. Such a representation can be effectively obtained by
introducing a subspace of the vector space $\limfunc{Span}\left\{ v,\gamma
_{0}\right\} $ and the corresponding orthogonal decomposition as in \cite[p.
10]{Pauli RT}, \cite[1]{Hes-74}%
\begin{equation}
x=x_{\Vert }+x_{\bot },\quad \text{where }x_{\Vert }\in \limfunc{Span}%
\left\{ v,\gamma _{0}\right\} \text{ and }x_{\bot }\text{ is orthogonal to }%
\limfunc{Span}\left\{ v,\gamma _{0}\right\} .  \label{xparxor1}
\end{equation}%
We will need also the corresponding relative velocity $\mathbf{v}$ which is
defined by 
\begin{gather}
\frac{\mathbf{v}}{\mathrm{c}}=\frac{v\wedge \gamma _{0}}{v\cdot \gamma _{0}}%
,\qquad v=\gamma \left( 1+\frac{\mathbf{v}}{\mathrm{c}}\right) \gamma
_{0}=\gamma \gamma _{0}\left( 1-\frac{\mathbf{v}}{\mathrm{c}}\right) ,
\label{xparxor2} \\
\text{where }\gamma =v\cdot \gamma _{0}=\left( 1-\frac{\mathbf{v}^{2}}{%
\mathrm{c}^{2}}\right) ^{-1/2}\text{ is the \emph{Lorentz factor}.}  \notag
\end{gather}%
Then we obtain the following identities:%
\begin{equation}
x\cdot v=\gamma \left[ x_{0}-\mathbf{x}_{\Vert }\cdot \frac{\mathbf{v}}{%
\mathrm{c}}\right] \text{ where }\mathbf{x}_{\Vert }=x_{\Vert }\wedge \gamma
_{0},  \label{xparxor3}
\end{equation}%
\begin{equation}
\sqrt{\left( x\cdot v\right) ^{2}-x^{2}}=\left\vert \gamma \left( \mathbf{x}%
_{\Vert }-x_{0}\frac{\mathbf{v}}{\mathrm{c}}\right) +\mathbf{x}_{\bot
}\right\vert \text{, where }\mathbf{x}_{\bot }=x_{\bot }\wedge \gamma _{0}.
\label{xparxor4}
\end{equation}%
Observe that the right-hand sides correspond to standard Lorentz boost
transformations for respectively time and space components of the vector $x$%
, \cite[p. 10]{Pauli RT}.

Consequently, we get the following representation of the scalar solution (%
\ref{KGfreg4b}) in the frame of an arbitrary observer $\gamma _{0}$:%
\begin{equation}
\psi \left( x\right) =\exp \left\{ \mp \mathrm{i}\kappa _{0}\gamma \left[
x_{0}-\mathbf{x}_{\Vert }\cdot \frac{\mathbf{v}}{\mathrm{c}}\right] \right\} 
\mathring{u}\left( \left\vert \gamma \left( \mathbf{x}_{\Vert }-x_{0}\frac{%
\mathbf{v}}{\mathrm{c}}\right) +\mathbf{x}_{\bot }\right\vert \right) .
\label{xparxor5}
\end{equation}%
The charge and current densities in the scalar case are given by the
expressions%
\begin{equation}
\rho =-\frac{\chi q}{m\mathrm{c}^{2}}\func{Im}\frac{\partial _{t}\psi }{\psi 
}\left\vert \psi \right\vert ^{2},\qquad \mathbf{J}=\frac{\chi q}{m}\func{Im}%
\frac{\nabla \psi }{\psi }\left\vert \psi \right\vert ^{2},  \label{xparxor6}
\end{equation}%
implying for the solutions $\psi _{\pm }$ in (\ref{KGfreg4}) the following
representation for the total conserved charge 
\begin{equation}
q_{\pm }=\int_{\mathbb{R}^{3}}\rho _{\pm }\left( t,\mathbf{x}\right) \,%
\mathrm{d}\mathbf{x}=\pm \frac{\chi q\omega _{0}}{m\mathrm{c}^{2}}=\pm q.
\label{KGfreg5}
\end{equation}%
The conserved energy $\mathcal{E}$ and momentum $\mathbf{p}$ densities are 
\begin{equation}
\mathcal{E}=\frac{\chi ^{2}}{2m}\left[ \frac{1}{\mathrm{c}^{2}}\tilde{%
\partial}_{t}\psi \tilde{\partial}_{t}^{\ast }\psi ^{\ast }+\tilde{\nabla}%
\psi \tilde{\nabla}^{\ast }\psi ^{\ast }+G\left( \psi ^{\ast }\psi \right)
+\kappa _{0}^{2}\psi \psi ^{\ast }\right] ,  \label{KGfreg6}
\end{equation}%
\begin{equation}
\mathbf{p}=\left( p^{1},p^{2},p^{3}\right) =-\frac{\chi ^{2}}{2m\mathrm{c}%
^{2}}\left( \tilde{\partial}_{t}\psi \tilde{\nabla}^{\ast }\psi ^{\ast }+%
\tilde{\partial}_{t}^{\ast }\psi ^{\ast }\tilde{\nabla}\psi \right) .
\label{KGfreg7}
\end{equation}%
In particular, the expression (\ref{KGfreg6}) for the energy density $%
\mathcal{E}$ implies the following representation for the total conserved
energy $\mathsf{E}_{\pm }$ for the wave function $\psi _{\pm }$ defined by (%
\ref{KGfreg4b}) 
\begin{equation}
\mathsf{E}_{\pm }=\chi \omega _{0}\left( 1+\frac{a_{\mathrm{C}}^{2}}{2a^{2}}%
\right) >0.  \label{KGfreg8}
\end{equation}%
A very detailed theory of the scalar Klein-Gordon equation including the
Lagrangian treatment can be found in \cite[1.5]{GreinerRQM}, \cite[1.1, p. 20%
]{Wachter}. In particular one can find there studies of EnMT $T^{\mu \nu }$
showing \emph{that the energy of solutions for both the positive and
negative frequencies is always positive}.

\subsection{Solutions to the spinor field equation}

We seek a solution to the free charge spinor equation (\ref{KGfreg1}) which
is expected to incorporate the features of the scalar solution as in (\ref%
{KGfreg4b}) and the plane-wave solution (\ref{frdira2}) to the Dirac
equation. We find that such a spinor solution does exist and is of the form $%
\psi \left( x\right) =\psi _{\mp }\left( v,x\right) $ 
\begin{equation}
\psi _{\mp }\left( v,x\right) =\psi _{0}\mathrm{e}^{\mp \mathrm{I}\mathbf{%
\sigma }_{3}\kappa _{0}x\cdot v}\mathring{u}\left( \sqrt{\left( x\cdot
v\right) ^{2}-x^{2}}\right) ,\qquad \kappa _{0}=\frac{\omega _{0}}{\mathrm{c}%
},  \label{KGfreg9}
\end{equation}%
where the Gaussian factor $\mathring{u}$ is defined by (\ref{KGfreg4a}) and $%
\psi _{0}$ is a normalized constant spinor from the even subalgebra $\mathrm{%
Cl}_{+}\left( 1,3\right) $ satisfying the canonical representation (\ref%
{frdira6}) and the following special conditions 
\begin{equation}
\left\langle \psi _{0}\tilde{\psi}_{0}\right\rangle =1\text{ or }%
\left\langle \psi _{0}\tilde{\psi}_{0}\right\rangle =-1,  \label{psinor1}
\end{equation}%
that is $\beta _{0}=0$ or $\beta _{0}=\pi $. One can readily see a distinct
feature of the spinor solution (\ref{KGfreg9}) compared to the plane-wave
solution (\ref{frdira2}) to the Dirac equation. It is the amplitude factor $%
\mathring{u}\left( \sqrt{\left( x\cdot v\right) ^{2}-x^{2}}\right) $ which
can be attributed to the nonlinearity $G\left( \left\langle \psi \tilde{\psi}%
\right\rangle \right) $ in the spinor equation (\ref{KGfreg1}). The origin
of the special constraints (\ref{psinor1}) can be traced to the particular
way $\psi $ enters the nonlinearity, namely as $G\left( \left\langle \psi 
\tilde{\psi}\right\rangle \right) $. For $\psi $ of the form (\ref{KGfreg9})
to be a solution to the free charge spinor equation (\ref{KGfreg1}), there
has to be an effective reduction to the scalar equation (\ref{KGfreg2}) with
the nonlinearity $G^{\prime }\left( \mathring{u}^{2}\right) $. The
constraint (\ref{psinor1}) is essential for such a reduction. Indeed, if the
spinor wave function $\psi $ is defined by (\ref{KGfreg9}) and satisfies the
condition (\ref{psinor1}) then%
\begin{equation}
\left\langle \psi \tilde{\psi}\right\rangle =\left\langle \psi _{0}\tilde{%
\psi}_{0}\right\rangle \mathring{u}^{2}\text{ implying }G^{\prime }\left(
\left\langle \psi \tilde{\psi}\right\rangle \right) =G^{\prime }\left( \pm 
\mathring{u}^{2}\right) =G^{\prime }\left( \mathring{u}^{2}\right) .
\label{psinor2}
\end{equation}%
When establishing identities (\ref{psinor2}) we used the identity (\ref%
{psinor3}) and that $G^{\prime }\left( s\right) $ defined by (\ref{nelagG2})
is an even function. The identities (\ref{psinor2}) allow to reduce the
spinor equation (\ref{KGfreg1}) to the scalar equation (\ref{KGfreg2}).

\subsection{Charge and current densities}

Let us consider the solution $\psi _{\mp }\left( x\right) $ as in (\ref%
{KGfreg9}) with $v=\gamma _{0}$, that is 
\begin{gather}
\psi _{\mp }\left( x\right) =\psi _{\mp }\left( \gamma _{0},x\right) =\psi
_{0}\mathrm{e}^{\mp \mathrm{I}\mathbf{\sigma }_{3}\kappa _{0}x_{0}}u\left(
x\right) ,\text{ where}  \label{psiuksi1} \\
x_{0}=x\cdot \gamma _{0},\qquad u\left( x\right) =\mathring{u}\left( \left(
x\cdot \gamma _{0}\right) ^{2}-x^{2}\right) ,  \notag
\end{gather}%
where the Gaussian factor $\mathring{u}$ is defined by (\ref{KGfreg4a}) and $%
\psi _{0}$ satisfies the condition (\ref{psinor1}), that is $\left\langle
\psi _{0}\tilde{\psi}_{0}\right\rangle =\pm 1$. Then the current component $%
J^{\mu }$ defined by (\ref{Jpiem2}) for the free charge with $\breve{A}=0$
takes the following form:%
\begin{equation}
J^{\mu }=\frac{q}{m}\left\langle \gamma ^{\mu }\chi \partial \psi \mathrm{I}%
\mathbf{\sigma }_{3}\tilde{\psi}\right\rangle =\frac{q}{m}\left\langle
\gamma ^{\mu }\chi \partial \psi \mathrm{I}\mathbf{\sigma }_{3}\tilde{\psi}%
\right\rangle .  \label{psiuksi3}
\end{equation}%
Since for the rest solution $\psi _{\mp }$ defined by (\ref{psiuksi1}) $%
\partial _{0}u=0$, the following relations hold:%
\begin{gather}
\partial _{0}\psi _{\mp }=\psi _{0}\left( \mp \mathrm{I}\mathbf{\sigma }%
_{3}\kappa _{0}\right) \mathrm{e}^{\mp \mathrm{I}\mathbf{\sigma }_{3}\kappa
_{0}x_{0}}u=\psi _{0}\mathrm{e}^{\mp \mathrm{I}\mathbf{\sigma }_{3}\kappa
_{0}x_{0}}\left( \mp \mathrm{I}\mathbf{\sigma }_{3}\kappa _{0}\right) u=\psi
_{\mp }\left( \mp \mathrm{I}\mathbf{\sigma }_{3}\kappa _{0}\right) u,
\label{psiuksi3a} \\
\partial _{j}\psi _{\mp }=\psi _{0}\mathrm{e}^{\mp \mathrm{I}\mathbf{\sigma }%
_{3}\kappa _{0}x_{0}}\partial _{j}u=\psi _{\mp }\partial _{j}\ln u,  \notag
\end{gather}%
\begin{equation}
\partial \psi _{\mp }=\gamma ^{\mu }\partial _{\mu }\psi _{0}\mathrm{e}^{\mp 
\mathrm{I}\mathbf{\sigma }_{3}\kappa _{0}x_{0}}u=\left[ \gamma ^{0}\psi
_{0}\left( \mp \mathrm{I}\mathbf{\sigma }_{3}\kappa _{0}\right)
u+\dsum_{1\leq j\leq 3}\gamma ^{j}\psi _{0}\partial _{j}u\right] \mathrm{e}%
^{\mp \mathrm{I}\mathbf{\sigma }_{3}\kappa _{0}x_{0}}.  \label{psiuksi4}
\end{equation}%
Notice that in view of the identity (\ref{psinor3}) we have%
\begin{equation}
\tilde{\psi}_{\mp }=\mathrm{e}^{\pm \mathrm{I}\mathbf{\sigma }_{3}\kappa
_{0}x_{0}}\tilde{\psi}_{0}u\left( x\right) .  \label{psiuksi5}
\end{equation}%
The above relation combined with the equality (\ref{psiuksi4}) yields%
\begin{gather}
\left\langle \gamma ^{0}\partial \psi _{\mp }\mathrm{I}\mathbf{\sigma }_{3}%
\tilde{\psi}_{\mp }\right\rangle =\left\langle \psi _{0}\left( \mp \mathrm{I}%
\mathbf{\sigma }_{3}\kappa _{0}\right) \mathrm{e}^{\mp \mathrm{I}\mathbf{%
\sigma }_{3}\kappa _{0}x_{0}}u\mathrm{I}\mathbf{\sigma }_{3}\mathrm{e}^{\pm 
\mathrm{I}\mathbf{\sigma }_{3}\kappa _{0}x_{0}}\tilde{\psi}%
_{0}u\right\rangle +  \label{psiuksi6} \\
+\dsum_{1\leq j\leq 3}u\partial _{j}u\left\langle \gamma ^{0}\gamma ^{j}\psi
_{0}\mathrm{I}\mathbf{\sigma }_{3}\tilde{\psi}_{0}\right\rangle  \notag \\
=\pm u^{2}\kappa _{0}\left\langle \psi _{0}\tilde{\psi}_{0}\right\rangle +%
\frac{1}{2}\dsum_{1\leq j\leq 3}\left( \partial _{j}u^{2}\right)
\left\langle \gamma ^{0}\gamma ^{j}\psi _{0}\mathrm{I}\mathbf{\sigma }_{3}%
\tilde{\psi}_{0}\right\rangle .  \notag
\end{gather}%
Hence, based on (\ref{psiuksi4}) and the above equality, we obtain%
\begin{gather}
J_{\mp }^{0}=\mathrm{c}\rho _{\mp }=\frac{\chi q}{m}\left\langle \gamma
^{0}\left( \partial \psi _{\mp }\right) \mathrm{I}\mathbf{\sigma }_{3}\tilde{%
\psi}_{\mp }\right\rangle =  \label{psiuksi7} \\
=\frac{\chi q}{m}\left[ \pm u^{2}\kappa _{0}\left\langle \psi _{0}\tilde{\psi%
}_{0}\right\rangle +\dsum_{1\leq j\leq 3}\frac{1}{2}\left( \partial
_{j}u^{2}\right) \left\langle \gamma ^{0}\gamma ^{j}\psi _{0}\mathrm{I}%
\mathbf{\sigma }_{3}\tilde{\psi}_{0}\right\rangle \right] ,  \notag
\end{gather}%
implying the following expressions for the total charge%
\begin{equation}
\mathsf{q}_{\mp }=\dint \rho _{\mp }\,\mathrm{d}\mathbf{x}=\pm \frac{\chi
\kappa _{0}}{m\mathrm{c}}q\left\langle \psi _{0}\tilde{\psi}%
_{0}\right\rangle =\pm q\left\langle \psi _{0}\tilde{\psi}_{0}\right\rangle .
\label{psiuksi8}
\end{equation}%
Observe that the different signs $\pm $ of the charge $\mathsf{q}_{\mp }$
above can be traced to the different signs of the frequencies in expressions
(\ref{psiuksi1}) for $\psi _{\mp }\left( x\right) $.

\subsection{Energy-momentum density}

Let us find the energy-momentum density for the solutions of the spinorial
field equations $\psi _{\mp }\left( x\right) $ defined by (\ref{psiuksi1}).
To facilitate efficient computation, we use the following identities.
Suppose $p_{\alpha }$, $c_{\alpha }$ and $\varphi $ are multivectors
satisfying%
\begin{equation}
p_{\alpha }=\varphi c_{\alpha },\qquad c_{\alpha }c_{\beta }=c_{\beta
}c_{\alpha },\qquad \tilde{c}_{\alpha }=c_{\alpha }.  \label{pafac1}
\end{equation}%
Then%
\begin{equation}
p_{\alpha }\tilde{p}_{\beta }=p_{\beta }\tilde{p}_{\alpha }=\varphi
c_{\alpha }c_{\beta }\tilde{\varphi}.  \label{pafac2}
\end{equation}%
Observe now that if we consider the derivatives $\partial _{\alpha }\psi
_{\mp }$ defined by (\ref{psiuksi3a}) and set%
\begin{equation}
p_{\alpha }=\partial _{\alpha }\psi _{\mp },\qquad c_{0}=\mp \mathrm{I}%
\mathbf{\sigma }_{3}\kappa _{0}u,\qquad c_{j}=\partial _{j}\ln u,\qquad
\varphi =\psi _{\mp },  \label{pafac2a}
\end{equation}%
then the relations (\ref{pafac2}) are satisfied, that is%
\begin{equation}
\left( \partial _{\alpha }\psi _{\mp }\right) \left( \partial _{\beta }%
\tilde{\psi}_{\mp }\right) =\left( \partial _{\beta }\psi _{\mp }\right)
\left( \partial _{\alpha }\tilde{\psi}_{\mp }\right) .  \label{pafac3}
\end{equation}%
Notice that the following relations hold for the solutions $\psi _{\mp }$
defined by (\ref{psiuksi1})%
\begin{equation}
\left\langle \psi _{\mp }\tilde{\psi}_{\mp }\right\rangle =\left\langle \psi
_{0}\tilde{\psi}_{0}\right\rangle u^{2},  \label{enmomps1}
\end{equation}%
\begin{equation}
\partial _{0}\psi _{\mp }=\mp \psi _{0}\mathrm{I}\mathbf{\sigma }_{3}\kappa
_{0}\mathrm{e}^{\mp \mathrm{I}\mathbf{\sigma }_{3}\kappa _{0}x_{0}}u,\qquad
\partial _{0}\tilde{\psi}_{\mp }=\pm u\mathrm{I}\mathbf{\sigma }_{3}\kappa
_{0}\mathrm{e}^{\pm \mathrm{I}\mathbf{\sigma }_{3}\kappa _{0}x_{0}}\tilde{%
\psi}_{0},  \label{enmomps2}
\end{equation}%
\begin{equation}
\left( \partial _{0}\psi _{\mp }\right) \left( \partial _{0}\tilde{\psi}%
_{\mp }\right) =-u^{2}\psi _{0}\mathrm{I}\mathbf{\sigma }_{3}\kappa _{0}%
\mathrm{e}^{\mp \mathrm{I}\mathbf{\sigma }_{3}\kappa _{0}x_{0}}\mathrm{I}%
\mathbf{\sigma }_{3}\kappa _{0}\mathrm{e}^{\pm \mathrm{I}\mathbf{\sigma }%
_{3}\kappa _{0}x_{0}}\tilde{\psi}_{0}=u^{2}\kappa _{0}^{2}\psi _{0}\tilde{%
\psi}_{0},  \label{enmomps3}
\end{equation}%
\begin{equation}
\partial _{j}\psi _{\mp }=\psi _{0}\mathrm{e}^{\mp \mathrm{I}\mathbf{\sigma }%
_{3}\kappa _{0}x_{0}}\partial _{j}u,\qquad \partial _{j}\tilde{\psi}_{\mp }=%
\mathrm{e}^{\pm \mathrm{I}\mathbf{\sigma }_{3}\kappa _{0}x_{0}}\tilde{\psi}%
_{0}\partial _{j}u,  \label{enmomps4}
\end{equation}%
\begin{equation}
\left( \partial _{j}\psi _{\mp }\right) \left( \partial _{j}\tilde{\psi}%
_{\mp }\right) =\psi _{0}\tilde{\psi}_{0}\left( \partial _{j}u\right) ^{2}.
\label{enmomps5}
\end{equation}%
Using then the expression (\ref{KGfreg0}) for the Lagrangian $L$ and the
identities (\ref{pafac3}), we find its value on the filed fields $\psi _{\mp
}$ to be%
\begin{gather}
L=\frac{\chi ^{2}}{2m}\left\{ \left\langle \gamma ^{\beta }\gamma ^{\alpha
}\left( \partial _{\alpha }\psi _{\mp }\right) \left( \partial _{\beta }%
\tilde{\psi}_{\mp }\right) \right\rangle -\left[ \kappa _{0}^{2}\left\langle
\psi _{\mp }\tilde{\psi}_{\mp }\right\rangle +G\left( \left\langle \psi
_{\mp }\tilde{\psi}_{\mp }\right\rangle \right) \right] \right\} =
\label{pafac4} \\
=\frac{\chi ^{2}}{2m}\left\{ \left\langle \left( \partial ^{\alpha }\psi
_{\mp }\right) \left( \partial _{\alpha }\tilde{\psi}_{\mp }\right)
\right\rangle -\left[ \kappa _{0}^{2}\left\langle \psi _{\mp }\tilde{\psi}%
_{\mp }\right\rangle +G\left( \left\langle \psi _{\mp }\tilde{\psi}_{\mp
}\right\rangle \right) \right] \right\} .  \notag
\end{gather}%
The canonical EnMT $\mathring{T}^{\mu \nu }$ defined by (\ref{Tmunp2}) takes
the following form for the free charge with $\breve{A}=0$ 
\begin{eqnarray}
\mathring{T}^{\mu \nu } &=&\frac{\chi }{m}\left\langle \gamma ^{\mu
}\partial ^{\nu }\psi \mathrm{I}\mathbf{\sigma }_{3}\left( \chi \partial
\psi \mathrm{I}\mathbf{\sigma }_{3}\right) ^{\symbol{126}}\right\rangle
-\delta _{\nu }^{\mu }L=\frac{\chi ^{2}}{m}\left\langle \gamma ^{\mu
}\partial ^{\nu }\psi \left( \partial \psi \right) ^{\symbol{126}%
}\right\rangle -\delta _{\nu }^{\mu }L=  \label{pafac5} \\
&=&\frac{\chi ^{2}}{m}\left\langle \gamma ^{\mu }\partial ^{\nu }\psi
\partial _{\alpha }\tilde{\psi}\gamma ^{\alpha }\right\rangle -\delta _{\nu
}^{\mu }L=\frac{\chi ^{2}}{m}\left\langle \gamma ^{\alpha }\gamma ^{\mu
}\partial ^{\nu }\psi \partial _{\alpha }\tilde{\psi}\right\rangle -\delta
_{\nu }^{\mu }L,  \notag
\end{eqnarray}%
where we used the identity (\ref{psinor3}). The above formula yields the
following representation for the energy density $\mathcal{E}$%
\begin{equation}
\mathcal{E}=\mathring{T}^{00}=\frac{\chi ^{2}}{m}\left\langle \gamma
^{\alpha }\gamma ^{0}\partial ^{0}\psi \partial _{\alpha }\tilde{\psi}%
\right\rangle -L.  \label{pafac6}
\end{equation}%
In particular, for $\psi =\psi _{\mp }$, we use (\ref{pafac6}) and (\ref%
{pafac4}) to obtain 
\begin{gather}
\mathcal{E}_{\mp }=\frac{\chi ^{2}}{m}\left\langle \gamma ^{\alpha }\gamma
^{0}\partial ^{0}\psi _{\mp }\partial _{\alpha }\tilde{\psi}_{\mp
}\right\rangle -L=  \label{pafac7} \\
=\frac{\chi ^{2}}{m}\left\langle \gamma ^{\alpha }\gamma ^{0}\partial
^{0}\psi _{\mp }\partial _{\alpha }\tilde{\psi}_{\mp }\right\rangle -\frac{%
\chi ^{2}}{2m}\left\langle \left( \partial ^{\alpha }\psi _{\mp }\right)
\left( \partial _{\alpha }\tilde{\psi}_{\mp }\right) \right\rangle +\frac{%
\chi ^{2}}{2m}\left[ \kappa _{0}^{2}\left\langle \psi _{\mp }\tilde{\psi}%
_{\mp }\right\rangle +G\left( \left\langle \psi _{\mp }\tilde{\psi}_{\mp
}\right\rangle \right) \right] .  \notag
\end{gather}%
The expression above can be transformed into%
\begin{gather}
\mathcal{E}_{\mp }=\frac{\chi ^{2}}{2m}\left\{ \left\langle \left( \partial
_{0}\psi _{\mp }\right) \left( \partial _{0}\tilde{\psi}_{\mp }\right)
\right\rangle +\dsum_{1\leq j\leq 3}\left\langle \left( \partial _{j}\psi
_{\mp }\right) \left( \partial _{j}\tilde{\psi}_{\mp }\right) \right\rangle +%
\left[ \kappa _{0}^{2}\left\langle \psi \tilde{\psi}\right\rangle +G\left(
\left\langle \psi \tilde{\psi}\right\rangle \right) \right] \right\} -
\label{pafac8} \\
-\dsum_{1\leq j\leq 3}\frac{\chi ^{2}}{m}\left\langle \gamma ^{j}\gamma
^{0}\partial ^{0}\psi _{\mp }\partial _{j}\tilde{\psi}_{\mp }\right\rangle .
\notag
\end{gather}%
Using the identities (\ref{enmomps1})-(\ref{enmomps5}) we transform the
above representation further into%
\begin{equation}
\mathcal{E}_{\mp }=\frac{\chi ^{2}\left\langle \psi _{0}\tilde{\psi}%
_{0}\right\rangle }{2m}\left[ 2u^{2}\kappa _{0}^{2}+\dsum_{1\leq j\leq
3}\left( \partial _{j}u\right) ^{2}+G\left( u^{2}\right) \right] \pm \frac{%
\chi ^{2}}{2m}\kappa _{0}\dsum_{1\leq j\leq 3}\left\langle \gamma ^{j}\gamma
^{0}\psi _{0}\mathrm{I}\mathbf{\sigma }_{3}\tilde{\psi}_{0}\right\rangle
\partial _{j}u^{2}.  \label{pafac9}
\end{equation}%
Then, using the above formula and relations (\ref{KGfreg4c}), (\ref{nelagG3}%
), we obtain the following representation for the total energy $\mathsf{E}%
_{\mp }$ of the free charge solutions $\psi _{\mp }$:%
\begin{eqnarray}
\mathsf{E}_{\mp } &=&\int \mathcal{E}_{\mp }\,\mathrm{d}\mathbf{x}=\frac{%
\chi ^{2}\left\langle \psi _{0}\tilde{\psi}_{0}\right\rangle }{2m}\int \left[
2u^{2}\kappa _{0}^{2}-\dsum_{1\leq j\leq 3}\left( \partial _{j}^{2}u\right)
u+G\left( u^{2}\right) \right] \,\mathrm{d}\mathbf{x}=  \label{pafac10} \\
&=&\frac{\chi ^{2}\left\langle \psi _{0}\tilde{\psi}_{0}\right\rangle }{2m}%
\int \left[ 2u^{2}\kappa _{0}^{2}-G^{\prime }\left( u^{2}\right)
u^{2}+G\left( u^{2}\right) \right] \,\mathrm{d}\mathbf{x}=  \notag \\
&=&\frac{\chi ^{2}\left\langle \psi _{0}\tilde{\psi}_{0}\right\rangle }{2m}%
\int \left( 2\kappa _{0}^{2}+\frac{1}{a^{2}}\right) u^{2}\,\mathrm{d}\mathbf{%
x}=\frac{\chi ^{2}\left\langle \psi _{0}\tilde{\psi}_{0}\right\rangle }{2m}%
\left( 2\kappa _{0}^{2}+\frac{1}{a^{2}}\right) =  \notag \\
&=&\left\langle \psi _{0}\tilde{\psi}_{0}\right\rangle \chi \omega
_{0}\left( 1+\frac{\mathrm{c}^{2}}{2\omega _{0}^{2}a^{2}}\right)
=\left\langle \psi _{0}\tilde{\psi}_{0}\right\rangle \chi \omega _{0}\left(
1+\frac{a_{\mathrm{C}}^{2}}{2a^{2}}\right) ,\qquad a_{\mathrm{C}}=\kappa
_{0}^{-1}=\frac{\chi }{m\mathrm{c}}.  \notag
\end{eqnarray}%
Observe now that if we want the energy $\mathsf{E}_{\pm }$ defined by (\ref%
{pafac10}) to be positive, and we do, then according to the canonical spinor
representation (\ref{frdira6}) for $\psi _{0}$ and constraints (\ref{psinor1}%
) we require $\left\langle \psi _{0}\tilde{\psi}_{0}\right\rangle =\cos
\beta _{0}=1$, that is $\beta _{0}=0$. This requirement in view of (\ref%
{frdira6}) is equivalent to the following constraint for free charge
solutions%
\begin{equation}
\psi _{0}\tilde{\psi}_{0}=1,\text{ that is }\psi _{0}\text{ is the Lorentz
rotor.}  \label{neopsiq5}
\end{equation}%
Consequently, under the above constraint the formula (\ref{pafac10}) turns
into 
\begin{equation}
\mathsf{E}_{\mp }=\chi \omega _{0}\left( 1+\frac{a_{\mathrm{C}}^{2}}{2a^{2}}%
\right) ,\qquad a_{\mathrm{C}}=\kappa _{0}^{-1}=\frac{\chi }{m\mathrm{c}}.
\label{pafac10a}
\end{equation}

Let us take a closer look at the origin of the factor $\left\langle \psi _{0}%
\tilde{\psi}_{0}\right\rangle $ in expressions (\ref{pafac10}) and (\ref%
{psiuksi8}) for $\mathsf{E}_{\mp }$ and $\mathsf{q}_{\mp }$. The similar
dependence on this factor of evidently spinorial nature occurs in the
quadratic part of the charge Lagrangian $L$ (without the nonlinear term $G$)
defined by (\ref{nelag1}). Observe that multiplication of the Lagrangian by
any constant positive or negative does not change the Euler-Lagrange
equation but it does alter the energy and the charge densities defined
canonically by the Lagrangian. \emph{To summarize, the presence of the
factor }$\left\langle \psi _{0}\tilde{\psi}_{0}\right\rangle $\emph{, which
can be positive or negative altering the sign of the Lagrangian, the energy
and the charge, is special to the spinorial wave functions since for
complex-valued ones the similar factor }$\psi _{0}\psi _{0}^{\ast }$\emph{\
is always positive. }

\section{Neoclassical solutions interpretation and comparison with the Dirac
theory\label{sneointer}}

As to general aspects of the interpretation of the wave function and
observables in the STA settings we rely mostly on works of D. Hestenes, see 
\cite[4]{Hes-75}, \cite[2]{Hes-96}, \cite[VII.D]{Hes2-03} and references
therein. The key of those aspects are as follows. First of all, based on the
general canonical representation (\ref{wfudisp1}) for the wave function $%
\psi =\psi \left( x\right) $, we assign at each spacetime point $x$ the 
\emph{local rotor} $R=R(x)$. This rotor determines the Lorentz rotation of a
given fixed frame of vectors $\{\gamma _{\mu }\}$ into the \emph{local rest
frame of vectors} $\{e_{\mu }=e_{\mu }(x)\}$ given by%
\begin{equation}
e_{\mu }=e_{\mu }(x)=R\gamma _{\mu }\tilde{R},\qquad R=R(x).
\label{psicanr3}
\end{equation}%
Importantly, in view of the canonical representation (\ref{wfudisp1}) we have%
\begin{equation}
\psi \gamma _{\mu }\tilde{\psi}=\varrho R\gamma _{\mu }\tilde{R}=\varrho
e_{\mu }.  \label{psicanr4}
\end{equation}%
The interpretation of the above fields in the Dirac theory is as follows.
The vector field%
\begin{equation}
\psi \gamma _{0}\tilde{\psi}=\varrho e_{0}=\varrho v  \label{psicanr4a}
\end{equation}%
is the \emph{Dirac current} (\emph{probability current} in the standard Born
interpretation) that determines the local rest frame $v$. \emph{The local
spin vector density} is defined by%
\begin{equation}
s=\frac{1}{2}\hbar \psi \gamma _{3}\tilde{\psi}=\frac{1}{2}\hbar \varrho
e_{3}.  \label{psicanr5}
\end{equation}%
The \emph{spin angular momentum} $S=S(x)$ (\emph{proper spin}) is a bivector
field related to the spin vector field $s=s(x)$ by, \cite[4]{Hes-75}, \cite[2%
]{Hes-96}%
\begin{equation}
S=\frac{1}{2}\hbar e_{2}e_{1}=\frac{1}{2}\hbar R\mathrm{I}\mathbf{\sigma }%
_{3}\tilde{R}=\frac{1}{2}\hbar R\gamma _{2}\gamma _{1}\tilde{R}=\frac{1}{2}%
\hbar R\mathrm{I}\mathbf{\sigma }_{3}\tilde{R}=\mathrm{I}se_{0}=\mathrm{I}%
\left( s\wedge e_{0}\right) .  \label{psicanr5a}
\end{equation}%
Notice that according to (\ref{Jpiem9}), (\ref{Jpiem10}) that the \emph{%
proper spin density} is $\varrho S$ and the \emph{magnetization} or \emph{%
magnetic moment density }$M$ of the charge is defined by the following
expression, 
\begin{equation}
M=\frac{\hbar e}{2m\mathrm{c}}\psi \gamma _{2}\gamma _{1}\tilde{\psi}=%
\mathrm{e}^{\mathrm{I}\beta }\frac{q}{m\mathrm{c}}\varrho S.
\label{psicanr6}
\end{equation}

Let us turn now to our neoclassical free charge solutions (\ref{psiuksi1})
satisfying energy positivity constraint (\ref{neopsiq5}), that is 
\begin{gather}
\psi _{\mp }\left( x\right) =\psi _{0}\mathrm{e}^{\mp \mathrm{I}\mathbf{%
\sigma }_{3}\kappa _{0}x_{0}}u\left( x\right) ,\text{ where }u\left(
x\right) =\mathring{u}\left( \left( x\cdot \gamma _{0}\right)
^{2}-x^{2}\right) ,\qquad \psi _{0}\tilde{\psi}_{0}=1,  \label{neopsiq1} \\
\mathring{u}\left( s\right) =a^{-3/2}\pi ^{-3/4}\exp \left( -\frac{s^{2}}{%
2a^{2}}\right) ,\qquad s\geq 0,  \label{neopsiq1a}
\end{gather}%
where the time-like vector unit vector $\gamma _{0}$ describes the constant
rest frame $v=\gamma _{0}$ of charge for every $x$. Then formulas (\ref%
{psiuksi8}) and (\ref{pafac10a}) yield the following ultimate expressions
for the total charge $\mathsf{q}_{\mp }$ and the total energy $\mathsf{E}%
_{\mp }$ of the solutions $\psi _{\mp }$ 
\begin{equation}
\mathsf{q}_{\mp }=\dint \rho _{\mp }\,\mathrm{d}\mathbf{x}=\pm q,\text{ for
the total charge,}  \label{neopsiq6}
\end{equation}%
\begin{equation}
\mathsf{E}_{\mp }=\int \mathcal{E}_{\mp }\,\mathrm{d}\mathbf{x}=\chi \omega
_{0}\left( 1+\frac{a_{\mathrm{C}}^{2}}{2a^{2}}\right) ,\text{ for the total
energy,}  \label{neopsiq7}
\end{equation}%
where%
\begin{equation}
\kappa _{0}=\frac{m\mathrm{c}}{\chi },\qquad \omega _{0}=\kappa _{0}\mathrm{c%
}=\frac{m\mathrm{c}^{2}}{\chi },\qquad a_{\mathrm{C}}=\kappa _{0}^{-1}=\frac{%
\chi }{m\mathrm{c}},  \label{neopsiq4}
\end{equation}%
and $\chi $ is a constant approximately equal to the Planck constant $\hbar $%
. Notice that subindices $\mp $ in $\mathsf{q}_{\mp }$ and $\mathsf{E}_{\mp
} $ are picked so that if the charge is electron then its index is "$-$" to
match the sign of the charge.

Observe that formula (\ref{neopsiq6}) demands the total charges $\mathsf{q}%
_{\mp }$ associated with the wave functions $\psi _{-}$ and $\psi _{+}$ to
have opposite signs. Following to the established tradition we call them 
\emph{charge and anticharge} (for instance, electron and positron). To see
"spinning" of the local rest frame as a defining basis for charge and
anticharge as two different states of the same single charge we introduce
the following representation of $\psi _{\mp }\left( x\right) $ based on (\ref%
{neopsiq1}) and (\ref{neopsiq5})%
\begin{equation}
\psi _{\mp }\left( x\right) =R_{\mp }\left( x\right) u\left( x\right) ,\quad 
\text{where }R_{\mp }=\psi _{0}\mathrm{e}^{\mp \mathrm{I}\mathbf{\sigma }%
_{3}\kappa _{0}x_{0}},\quad \psi _{0}\tilde{\psi}_{0}=1\qquad R_{\mp }\tilde{%
R}_{\mp }=1.  \label{neopsiq8}
\end{equation}%
Then following to D. Hestenes, \cite[6, 9]{Hes-81} and using the above
representation we obtain a formula involving the rotational velocities $%
\Omega _{\mp }$: 
\begin{equation}
\frac{d\psi _{\mp }}{dt}=\frac{\mathrm{c}d\psi _{\mp }}{dx_{0}}=\mathrm{c}%
\frac{1}{2}\Omega _{\mp }\psi _{\mp },\qquad \Omega _{\mp }=\mp 2\kappa _{0}%
\mathrm{c}R_{\mp }\left( \mathrm{I}\mathbf{\sigma }_{3}\right) \tilde{R}%
_{\mp }.  \label{neopsiq9}
\end{equation}%
In other words, \emph{the local rest frame rotates "about the spin axis" }$%
s=R_{\mp }\left( \gamma _{3}\right) \tilde{R}_{\mp }$\emph{\ with the
angular speed }$\left\vert \Omega _{\mp }\right\vert =2\kappa _{0}c=2\omega
_{0}$\emph{\ equal to twice the frequency }$\omega _{0}$\emph{\
corresponding the rest energy }$mc^{2}=\chi \omega _{0}$\emph{.} Observe
that according to relations (\ref{neopsiq9}) the \emph{charge and
"anticharge" (for instance, electron and positron) wave functions }$\psi
_{-} $\emph{\ and }$\psi _{+}$\emph{\ differ only by ascribing opposite
sense to the rotation described by the factor} $\mathrm{e}^{\mp \mathrm{I}%
\mathbf{\sigma }_{3}\kappa _{0}x_{0}}$ in (\ref{neopsiq8}).

Let consider now the magnetization $M$ and the proper spin $\varrho S$
bivector densities defined by (\ref{psicanr5a}),(\ref{psicanr6}) for the
neoclassical free charge solutions $\psi _{\mp }$ described by formulas (\ref%
{neopsiq1}) and (\ref{neopsiq8}). Notice that for the free charge at rest we
have $\varrho =u^{2}\left( x\right) $ and the following relation holds for
the magnetization and proper spin densities%
\begin{equation}
M_{\mp }=\frac{\hbar e}{2m\mathrm{c}}\psi _{\mp }\gamma _{2}\gamma _{1}%
\tilde{\psi}_{\mp }=\frac{q}{m\mathrm{c}}u^{2}\left( x\right) S_{\mp }\text{%
, where }S_{\mp }=\frac{1}{2}\hbar \psi _{0}\mathrm{I}\mathbf{\sigma }_{3}%
\tilde{\psi}_{0},  \label{magsmom1}
\end{equation}%
where we took into account that $\beta =0$ since $\psi _{0}\tilde{\psi}%
_{0}=1 $. Observe also that the above formula shows that the both states $%
\psi _{\mp }$ have the same and constant $M_{\mp }$ and $S_{\mp }$.
Integrating then the magnetization density of the free charge at rest $%
M_{\mp }$ in (\ref{magsmom1}) we obtain the total generalized magnetic
momentum bivector%
\begin{equation}
\mathcal{M}_{\mp }=\int_{\mathbb{R}^{3}}M_{\mp }\,\mathrm{d}\mathbf{x}=\frac{%
q}{m\mathrm{c}}S_{\mp }.  \label{magsmom2}
\end{equation}%
For general issues of the STA treatment of localized charge distributions
and the proper momentum bivector $M$ see \cite[1]{Hes2-74}.

Let us compare now the neoclassical free charge solutions (\ref{neopsiq1})
with the Dirac free charge solutions (\ref{frdira1}). First of all, the
spinorial aspect of the proposed here neoclassical theory is identical to
that in the Dirac theory since in the both cases the wave function $\psi $
takes values in the even algebra $\mathrm{Cl}_{+}\left( 1,3\right) $. The
governing field equation for the neoclassical spinor field is (\ref{KGfreg1}%
) and the Dirac spinor field satisfies the Dirac equation (\ref{dirac1}).
Structurally the neoclassical field equation (\ref{KGfreg1}) can be viewed
as a spinorial version of the Klein-Gordon equation with added nonlinearity
and consequently it is related to the Dirac equation. In particular, \emph{%
solutions to the Dirac equations are also solutions to our field equations
if the nonlinearity there is neglected}. But when it comes to the structure
of solutions the first significant difference of the neoclassical free
charge solution (\ref{neopsiq1}) compare to the Dirac free charge plane wave
solution is the Gaussian factor $u\left( x\right) $. In other words,\emph{\
the neoclassical free charge solution is a localized soliton-like wave
whereas the Dirac free charge solution is a plane wave}.

In what follows we compare other features of the neoclassical free charge
wave function and the Dirac free charge plane wave function.

\subsection{The gyromagnetic ratio and currents}

Recall that the gyromagnetic ratio $g$ is defined as a coefficient that
relates the magnetic dipole moment $\mathbf{m}$ and the angular momentum $%
\mathbf{L}$ for a system of localized currents, namely%
\begin{equation}
\mathbf{m}=\frac{gq}{2m\mathrm{c}}\mathbf{L}.  \label{mgyroL1}
\end{equation}%
Comparing the above relation (\ref{mgyroL1}) with (\ref{magsmom2}) we
conclude that in our theory the gyromagnetic ratio $g=2$ just as in the
Dirac theory. \emph{The expressions (\ref{Jpiem6})-(\ref{Jpiem10}) for the
current and its Gordon decomposition which includes the magnetization (spin)
current in the neoclassical are identical to the same in the Dirac theory} (%
\ref{ecurr4})-(\ref{ecurr7}). This identity of the above mentioned currents
is very important since they were extensively analyzed and thoroughly tested
experimentally. The value of the gyromagnetic ratio $g=2$ in our theory is\
in fact not so surprising since the minimal coupling as in (\ref{nelag2}) 
\emph{implies that }$g=2$, \cite[2-2-3]{ItzZub}. Interestingly, there is an
example of a \emph{classical particle with the gyromagnetic ratio} $g=2$, 
\cite[V]{Hes2-03}.

\subsection{The energies and frequencies}

The issue of negative energies in the Dirac theory constitutes a well known
serious problem discussed extensively in the literature, see for instance 
\cite[12]{GreinerRQM}, \cite[2.4.2]{ItzZub}, \cite[2.1.6]{Wachter} and
references therein. One of the proposed ways to deal with it is \emph{to
reinterpret an electron state of negative energy/frequency as a positron
state of positive energy/frequency using the charge conjugation
transformation (\ref{frdira14})}. This operation in the conventional setting
involves complex conjugation of the Dirac wave function and reverses the
sign of the charge, its frequency, energy, momentum, and spin, \cite[5.4]%
{HalzenMartin}, \cite[2.1.6, p. 109]{Wachter}. Effectively, the charge
conjugation operation changes the sign of the frequency of the wave function
which satisfies then a complex conjugate version of the original Dirac
equation with the opposite sign of the charge there.

In quantum mechanics and in the Dirac theory in particular, the energy is
identified\ with the frequency via the Planck-Einstein relation $\mathcal{E}%
=\hbar \omega $. \ In contrast, in our neoclassical theory the energy and
the frequency  are two distinct though closely related concepts, see \cite%
{BF7}. The relation between them in the relativistic case by no means is as
explicit as the Planck-Einstein relation $\mathcal{E}=\hbar \omega $.
Importantly, in our theory the frequencies may be positive or negative when
the energy is positive. The positivity of the energies of the free charge in
our theory was obtained by simply limiting the values of the spinor constant 
$\psi _{0}$ for the free charge solutions $\psi _{\mp }$ in (\ref{neopsiq1})
to be a Lorentz rotor, that is to satisfy the energy positivity constraint (%
\ref{neopsiq5}), i.e. $\psi _{0}\tilde{\psi}_{0}=1$. Consequenty, in our
theory a positron state differs from the quantum mechanical positron state
and, importantly, it is not obtained by applying the charge conjugation (\ref%
{frdira14}) to an electron state. Also there were no changes of frequencies
or any transformation of the evolution equation. If the energy positivity
constraint (\ref{neopsiq5}) is satisfied, then according to (\ref{neopsiq7})
the energies $\mathsf{E}_{\mp }$ of the free charge at rest satisfy
approximately $\mathsf{E}_{\mp }\approx \hbar \left\vert \omega
_{0}\right\vert $, and they stay positive whereas the corresponding
frequencies $\mp \kappa _{0}\mathrm{c}=\mp \omega _{0}$ of the solutions $%
\psi _{\mp }$ in (\ref{neopsiq1}) can be positive or negative. The above
analysis indicates a significant difference in the treatment of negative
energies in our theory compare to the Dirac theory or the QM. One may notice
though that we analyzed this far only energies of free charges. We expect
the treatment of a charge in external field to be more complex, but this
study is left for the future work.

\subsection{Antimatter states}

Similarly to the Dirac theory our theory naturally integrates into it the
concept of an antiparticle. According to (\ref{neopsiq8}) there are two
directions of "spinning" in the rest frame and that naturally leads\ to the
concepts of charge and anticharge with the frequencies of the opposite signs.%
\emph{\ }Note that since the value of a charge is preserved even in external
EM field the charge can not turn into the anti-charge as a result of
electromagnetic interactions. All the properties of the charge and
anticharge are exactly the same except for the difference in sign. We would
like stress once again a noticebale difference between the antimatter states
in our theory and the same in the Dirac theory. In our theory the matter and
antimatter states correspond to $\left\langle \psi _{0}\tilde{\psi}%
_{0}\right\rangle =1$, whereas in the Dirac theory the usual way to
introduce the antimatter state (positron) is by applying charge conjugation (%
\ref{frdira14}) that requires $\left\langle \psi _{0}\tilde{\psi}%
_{0}\right\rangle =\left\langle \mathrm{e}^{\beta _{0}\mathrm{I}%
}\right\rangle =\cos \beta _{0}=-1$ as in (\ref{frdira6}), (\ref{frdira12b}%
). Consequently, in order to introduce the antimatter state in the Dirac
theory one has to invoke the parameter $\beta $ of the canonical spinor
representation (\ref{wfudisp2}), but the interpretation of the parameter $%
\beta $ has known difficulties, \cite[3]{Hes-96}, \cite[VII.D, G]{Hes2-03}.

\textbf{Acknowledgment.} The research was supported through Dr. A. Nachman
of the U.S. Air Force Office of Scientific Research (AFOSR), under grant
number FA9550-11-1-0163.

\end{document}